\documentclass[11pt]{article}
\usepackage[left=1in,right=1in,top=1in]{geometry}

\usepackage{times}
\usepackage{amsmath,amssymb,amsthm,array,authblk,enumitem,hyperref,multirow,url,xspace}

\usepackage{graphicx}
\usepackage{color}
\usepackage[caption=false]{subfig}

\usepackage{algorithm}
\usepackage[noend]{algpseudocode}

\usepackage[sort,numbers]{natbib}

\newtheorem{definition}{Definition}
\newtheorem{example}{Example}
\newtheorem{lemma}{Lemma}
\newtheorem{theorem}{Theorem}

\DeclareMathOperator*{\argmin}{argmin}
\DeclareMathOperator*{\argmax}{argmax}

\newcommand{\rss}{RSS\xspace}
\newcommand{\sm}{SM\xspace}
\newcommand{\pName}{SMDK\xspace}

\newcommand{\algDen}{\textsc{CostEffectGreedy}\xspace}
\newcommand{\algStream}{\textsc{KnapStream}\xspace}
\newcommand{\histogram}{\textsc{SubKnapChk}\xspace}
\newcommand{\algWindow}{\textsc{KnapWindow}\xspace}
\newcommand{\algWindowOpt}{\textsc{KnapWindowPlus}\xspace}
\newcommand{\optimal}{\mathtt{OPT}}

\newcommand{\shortDen}{\text{CEG}}
\newcommand{\shortCardWin}{\text{WIN}}
\newcommand{\shortStr}{\text{STR}}
\newcommand{\shortKs}{\text{KS}}
\newcommand{\shortWin}{\text{KW}}
\newcommand{\shortWinOpt}{\text{KW}^{+}}

\begin{document}
\title{Efficient Representative Subset Selection over Sliding Windows}

\author[1]{Yanhao Wang}
\author[2]{Yuchen Li}
\author[1]{Kian-Lee Tan}
\affil[1]{School of Computing, National University of Singapore, Singapore}
\affil[2]{School of Information Systems, Singapore Management University, Singapore}
\affil[ ]{$^1$\textit{\{yanhao90, tankl\}@comp.nus.edu.sg\quad $^2$yuchenli@smu.edu.sg}}

\maketitle

\begin{abstract}
  Representative subset selection (RSS) is an important tool for users
  to draw insights from massive datasets. Existing literature models
  RSS as the submodular maximization problem to capture the ``diminishing
  returns'' property of the representativeness of selected subsets, but
  often only has a single constraint (e.g., cardinality), which limits its
  applications in many real-world problems. To capture the data recency
  issue and support different types of constraints, we formulate dynamic
  RSS in data streams as maximizing submodular functions subject to general
  $d$-knapsack constraints (SMDK) over sliding windows.
  We propose a \textsc{KnapWindow} framework (KW) for SMDK. KW
  utilizes the \textsc{KnapStream} algorithm (KS) for SMDK in
  append-only streams as a subroutine. It maintains a sequence of checkpoints and KS
  instances over the sliding window. Theoretically, KW is
  $\frac{1-\varepsilon}{1+d}$-approximate for SMDK.
  Furthermore, we propose a \textsc{KnapWindowPlus}
  framework (KW$^{+}$) to improve upon KW.
  KW$^{+}$ builds an index \textsc{SubKnapChk} to manage the checkpoints and KS instances.
  \textsc{SubKnapChk} deletes a checkpoint whenever it can be approximated by its successors.
  By keeping much fewer checkpoints, KW$^{+}$ achieves higher efficiency than KW
  while still guaranteeing a $\frac{1-\varepsilon'}{2+2d}$-approximate
  solution for SMDK. Finally, we evaluate the efficiency and solution quality
  of KW and KW$^{+}$ in real-world datasets. The experimental results demonstrate that
  KW achieves more than two orders of magnitude speedups over the batch baseline
  and preserves high-quality solutions for SMDK over sliding windows.
  KW$^{+}$ further runs 5-10 times faster than KW while providing solutions
  with equivalent or even better utilities.
\end{abstract}

\section{Introduction}
\label{sec:introduction}

In the big data era, a vast amount of data is being continuously generated
by various applications, e.g., social media, network traffic, sensors, etc.
An imperative task is to extract useful information from massive datasets.
A compelling approach is \emph{representative subset
selection}~\cite{DBLP:conf/icml/GomesK10,DBLP:conf/acl/LinB11,DBLP:conf/naacl/WeiLKB13,DBLP:conf/sigmod/XuKM14,DBLP:conf/kdd/BadanidiyuruMKK14,DBLP:conf/icml/WeiIB15,DBLP:conf/cikm/ZhuangRHGHA16,DBLP:conf/nips/LindgrenWD16,DBLP:conf/uai/MalioutovKY16,DBLP:conf/icml/MirzasoleimanBK16}
(\rss): extracting a concise subset of representative elements from the source dataset.
\rss is often formulated as selecting a subset of elements
to maximize a utility function that quantifies the \emph{representativeness}
subject to some \emph{constraints}.
The utility functions are often chosen to be \emph{submodular} to capture the ``diminishing returns'' property of representativeness~\cite{DBLP:conf/icml/GomesK10,DBLP:conf/kdd/BadanidiyuruMKK14,DBLP:conf/cikm/ZhuangRHGHA16,DBLP:conf/ijcai/TamWTYH17,DBLP:conf/globalsip/YuXC16,DBLP:journals/tkde/FanLZT17},
i.e., adding more elements decreases the marginal representativeness.
A number of constraints are used to restrict the selected subset in various ways.
For example, a common approach to scaling kernel methods in nonparametric learning is
\emph{active set selection}~\cite{DBLP:conf/icml/GomesK10,DBLP:conf/kdd/BadanidiyuruMKK14}
that extracts a subset $S$ with the maximum information entropy as representatives.
It restricts the size of $S$ to $k$ (a.k.a cardinality constraint)
so that at most $k$ elements are selected for kernel training
to reduce the computational costs while still retaining model quality.
As another example, \emph{social data summarization}~\cite{DBLP:conf/cikm/ZhuangRHGHA16,DBLP:conf/ijcai/TamWTYH17}
selects a subset $S$ to best preserve the information in a collection of social posts.
To restrict the summary size, two constraints are imposed:
the number of selected posts in $S$, as well as their total length,
is bounded. Additionally, the influence scores are also
modeled as constraints so that more influential elements could be included in the summary~\cite{DBLP:conf/globalsip/YuXC16}.

In many cases, data is generated rapidly and only available as a
stream~\cite{DBLP:conf/icde/LiZLT16,DBLP:journals/pvldb/GuoLST17,DBLP:journals/pvldb/ShaLHT17,DBLP:conf/icde/LiBLT15,DBLP:journals/tois/ZhangLFGSS17}.
To address the requirement for summarizing such datasets in real-time,
\rss over data streams~\cite{DBLP:conf/sdm/SahaG09,DBLP:conf/icml/GomesK10,DBLP:conf/kdd/BadanidiyuruMKK14,DBLP:conf/ijcai/TamWTYH17}
has been extensively studied in recent years.
However, there are two major drawbacks that limit the deployment of
existing approaches to many real-world applications.
First, most of the streaming \rss algorithms only work with cardinality constraints,
i.e., selecting a set of $k$ elements as representatives,
and cannot support more complex constraints.
As aforementioned, a number of \rss problems consider
more general multi-knapsack (a.k.a. $d$-knapsack) constraints beyond
cardinality~\cite{DBLP:conf/naacl/LinB10,DBLP:conf/acl/LinB11,DBLP:conf/globalsip/YuXC16,DBLP:conf/icml/MirzasoleimanBK16}.
However, the algorithms that only support cardinality constraints
cannot provide solutions with any quality assurances in more general cases.
Second, existing methods are developed for the append-only setting
where elements are only inserted into but never deleted from the stream
and thus the freshness of solutions is ignored.
Data streams are highly dynamic and keep evolving over time,
where recent elements are more important than earlier ones.
The sliding window~\cite{DBLP:journals/siamcomp/DatarGIM02} model
that only considers the $W$ most recent elements
is a natural way to capture such an essence.
Although a number of \rss algorithms have been developed for append-only streams,
\rss over sliding windows is still largely unexplored
and, to the best of our knowledge, only one existing method~\cite{DBLP:conf/www/EpastoLVZ17}
is proposed. It is not surprising that the method is also specific for cardinality constraints.

To address the limitations of existing methods,
it requires general \rss frameworks that
(i) support different types of submodular utility functions,
(ii) work with more than one knapsack constraint, and
(iii) extract a subset of representative elements over a sliding window efficiently.

In this paper, we formulate dynamic \rss in data streams as maximizing submodular functions
with $d$-knapsack constraints (\pName) over sliding windows.
As \pName is NP-hard, we focus on designing efficient approximation algorithms for \pName.
First, we devise the \algStream algorithm ($\shortKs$) for \pName
in append-only streams. $\shortKs$ needs a single pass
over a stream and provides a $\frac{1-\varepsilon}{1+d}$-approximate solution for \pName.
It improves the state-of-the-art approximation factor of
$\frac{1}{1+2d}-\varepsilon$ for \pName in append-only streams~\cite{DBLP:conf/globalsip/YuXC16}.
Then, we propose two novel frameworks, namely \algWindow ($\shortWin$) and \algWindowOpt ($\shortWinOpt$),
for \pName over sliding windows. Both frameworks adapt $\shortKs$ for the sliding window model
by maintaining a sequence of $\shortKs$ instances starting at different timestamps
(a.k.a \emph{checkpoints}) over the sliding window.
Specifically, $\shortWin$ maintains $\mathcal{O}(\sqrt{W})$ checkpoints
over a size-$W$ sliding window.
The interval between any neighboring checkpoints of $\shortWin$ is always equal.
The approximation factor of $\shortWin$ for \pName is the same as $\shortKs$,
i.e., $\frac{1-\varepsilon}{1+d}$.
Furthermore, $\shortWinOpt$ is proposed to build an index \histogram to manage
the checkpoints based on their achieved utilities.
\histogram deletes a checkpoint whenever it can be approximated by its successors.
Theoretically, the number of checkpoints in $\shortWinOpt$ is independent of $W$
and logarithmic to the range of the utility function.
Since $\shortWinOpt$ maintains much fewer checkpoints,
it achieves higher efficiency than $\shortWin$.
Nevertheless, $\shortWinOpt$ can still guarantee
$\frac{1-\varepsilon'}{2+2d}$-approximation solutions for \pName over sliding windows.

Finally, we evaluate the efficiency and effectiveness of $\shortWin$ and $\shortWinOpt$
with two real-world applications: \emph{social stream summarization}
and \emph{active set selection}. The experimental results show that
$\shortWin$ achieves more than two orders of magnitude speedup over the batch baseline
and preserves high-quality solutions for \pName over sliding windows.
$\shortWinOpt$ further runs 5--10 times faster than $\shortWin$
while providing solutions with equivalent or even better utilities.

Our main contributions are summarized as follows.
\begin{itemize}
  \item We formulate dynamic \rss as maximizing submodular functions
  with $d$-knapsack constraints (aka SMDK) over sliding windows.
  \item We propose a novel $\frac{1-\varepsilon}{1+d}$-approximation
  $\shortWin$ framework for \pName over sliding windows.
  \item We devise $\shortWinOpt$ to improve upon $\shortWin$.
  Although the approximation factor of $\shortWinOpt$ drops to $\frac{1-\varepsilon'}{2+2d}$,
  $\shortWinOpt$ has much higher efficiency than $\shortWin$ while providing solutions
  with equivalent or better quality.
  \item We demonstrate the efficiency and solution quality of $\shortWin$ and $\shortWinOpt$
  for real-world applications.
\end{itemize}

The remaining of this paper is organized as follows.
Section~\ref{sec:definition} defines dynamic \rss as \pName over sliding windows.
Section~\ref{sec:application} gives two examples of modeling real-world \rss
applications as \pName.
Section~\ref{sec:framework} and Section~\ref{sec:kwplus} present the $\shortWin$
and $\shortWinOpt$ frameworks respectively.
Section~\ref{sec:experiment} reports the experimental results.
Section~\ref{sec:related} reviews the related work.
Finally, Section~\ref{sec:conclusion} concludes the whole paper.

\section{Problem Formulation}\label{sec:definition}

In this section, we first introduce data streams and the sliding window model.
Next, we give the notions of \emph{submodular functions} and \emph{knapsack
constraints}. Then, we formally define the \emph{representative subset selection}
(\rss) problem as submodular maximization with a $d$-knapsack constraint (\pName)
in the sliding window model. Finally, we show the challenges of \pName over
sliding windows.

\textbf{Data Stream \& Sliding Window.}
A data stream comprises an unbounded sequence of elements $V=\langle v_1,v_2,\ldots \rangle$
and $v_t \in V$ is the $t$-th element of the stream. The elements in $V$ arrive
one at a time in an arbitrary order. Only \emph{one pass} over the
stream is permitted and the elements must be processed in the arrival order.
Specifically, we focus on the \emph{sliding window} model for data streams.
Let $W$ be the size of the sliding window. At any time $t$, the \emph{active window} $A_t$
is a subsequence that
always contains the $W$ most recent elements (a.k.a. \emph{active elements}) in the
stream\footnote{We only discuss the sequence-based sliding window in this
paper. Nevertheless, the proposed algorithms can naturally support the time-based sliding
window.}, i.e., $A_t=\langle v_{t'},\ldots,v_t \rangle$ where $t'=\max(1,t-W+1)$.

\textbf{\rss over Sliding Windows.}
\rss selects a set of representative elements from the ground set according to
a utility function with some budget constraint. In this paper, we target the
class of \emph{nonnegative monotone submodular} utility functions adopted in
a wide range of \rss problems~\cite{DBLP:conf/kdd/BadanidiyuruMKK14,DBLP:conf/www/EpastoLVZ17,DBLP:conf/icml/GomesK10,DBLP:conf/nips/LindgrenWD16,DBLP:conf/ijcai/TamWTYH17,DBLP:conf/sigmod/XuKM14,DBLP:conf/cikm/ZhuangRHGHA16}.

Given a ground set of elements $V$, we consider a set function
$f:2^V \rightarrow \mathbb{R}_{\geq 0}$ that maps
any subset of elements to a nonnegative \emph{utility} value. For a set of
elements $S \subseteq V$ and an element $v \in V \setminus S$, the \emph{marginal
gain} of $f(\cdot)$ is defined by $\Delta_f(v|S) \triangleq f(S\cup\{v\})-f(S)$.
Then, the \emph{monotonicity} and \emph{submodularity} of $f(\cdot)$ can be
defined according to its marginal gain.
\begin{definition}[Monotonicity \& Submodularity]
  A set function $f(\cdot)$ is monotone iff $\Delta_f(v|S) \geq 0$ for any
  $S \subseteq V$ and $v \in V \setminus S$. $f(\cdot)$ is submodular iff
  $\Delta_f(v|S) \geq \Delta_f(v|S')$ for any $S \subseteq S' \subseteq V$
  and $v \in V \setminus S'$.
\end{definition}
Intuitively, monotonicity means adding more elements does not decrease the
utility value. Submodularity captures the ``diminishing returns'' property that
the marginal gain of adding any new element decreases as a set grows larger.

To handle various types of \emph{linear} budget constraints in real-world
problems, we adopt the general $d$-knapsack constraint~\cite{DBLP:conf/naacl/LinB10,DBLP:conf/acl/LinB11,DBLP:conf/icml/MirzasoleimanBK16,DBLP:conf/globalsip/YuXC16}.
Specifically, a {\em knapsack} is defined by a cost function $c:V\rightarrow\mathbb{R}_{+}$
that assigns a positive cost to each element in the ground set $V$. Let $c(v)$
denote the cost of $v \in V$. The cost $c(S)$ of a set $S \subseteq V$ is
the sum of the costs of its members, i.e., $c(S)=\sum_{v \in S}c(v)$. Given a
budget $b$, we say $S$ satisfies the knapsack constraint iff $c(S) \leq b$.
W.l.o.g., we normalize the budget to $b=1$ and the cost of any element to
$c(v) \in (0,1]$. Then, a $d$-knapsack constraint $\xi$ is defined
by $d$ cost functions $c_1(\cdot),\ldots,c_d(\cdot)$. Formally, we define
$\xi = \{S \subseteq V : c_j(S) \leq 1, \forall j \in [d]\}$.
We say a set $S$ satisfies the $d$-knapsack constraint iff $S\in\xi$.

Given the above definitions, we can formulate \rss as an optimization problem
of maximizing a monotone submodular utility function $f(\cdot)$ subject to
a $d$-knapsack constraint $\xi$ (\pName) over the active window $A_t$.
At every time $t$, \rss returns a subset of elements $S_t$ that (1)
only contains active elements, (2) satisfies the $d$-knapsack
constraint $\xi$, and (3) maximizes the utility function $f(\cdot)$.
Formally,
\begin{equation}\label{eq:smdk}
  \max_{S_t \subseteq A_t} f(S_t) \quad \text{s.t.} \quad  S_t \in \xi
\end{equation}
We use $S^*_t=\argmax_{S_t \subseteq A_t : S_t \in \xi} f(S_t)$
to denote the optimal solution of \pName at time $t$.

\begin{figure}
  \centering
  \includegraphics[width=0.6\textwidth]{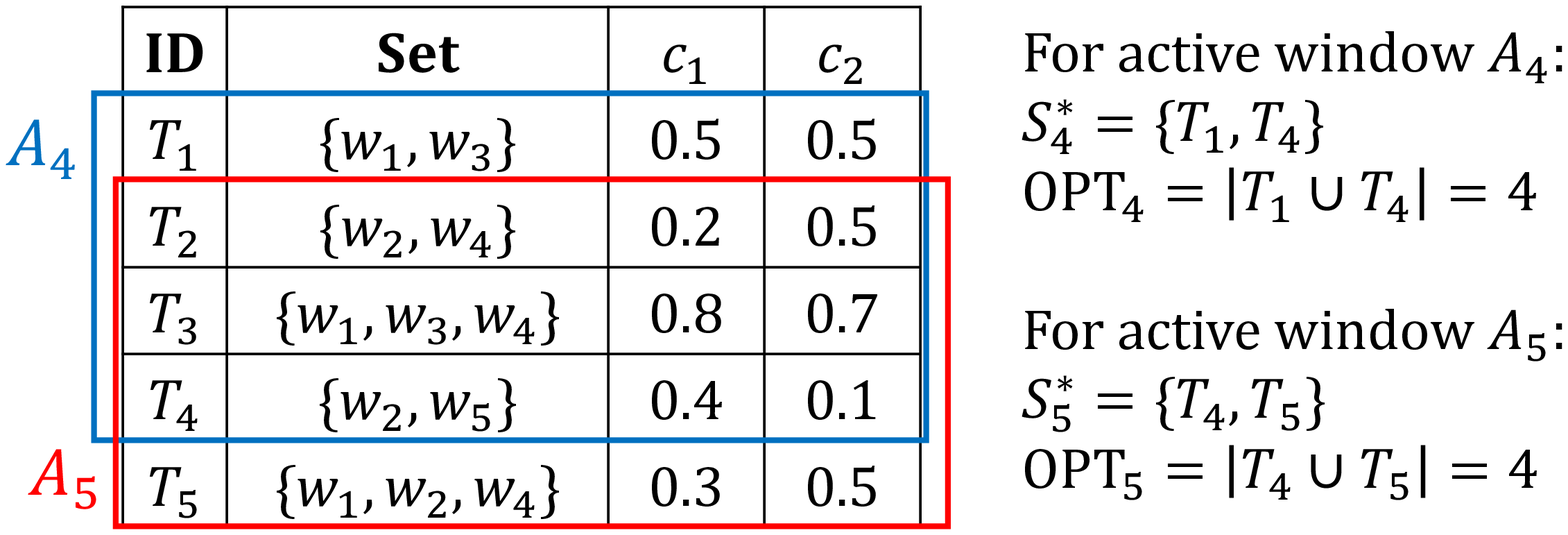}
  \caption{Toy example of \pName over sliding windows.
  We highlight two active windows $A_4,A_5$
  and show their optimal solutions and utilities.}
  \label{fig:toy:example}
\end{figure}

\begin{example}
  A toy example of \pName over sliding windows is given in Figure~\ref{fig:toy:example}.
  We consider one of the simplest \pName problems:
  budgeted maximum coverage (BMC)~\cite{DBLP:journals/ipl/KhullerMN99}.
  Given a domain of items $\mathcal{W}=\{w_1,\ldots,w_5\}$,
  we have a sequence of sets $\mathcal{T}=\langle T_1,\ldots,T_5 \rangle$
  where each set $T \in \mathcal{T}$ is a subset of $\mathcal{W}$
  associated with two costs $c_1$ and $c_2$.
  Let the window size be $4$.
  The objective of BMC is to select a set of sets $S^*_t$ from $4$ most recent sets
  such that the number of items covered by $S^*_t$ is maximized while $S^*_t$
  satisfies the $2$-knapsack constraint defined by $c_1$ and $c_2$.
  In Figure~\ref{fig:toy:example}, we highlight two active windows $A_4$ and $A_5$
  at time $4$ and $5$ respectively. Then, we give the optimal solutions $S^*_4$ and $S^*_5$
  and their utilities for BMC at time $4$ and $5$.
\end{example}

\textbf{Challenges of \pName over Sliding Windows.}
\pName is NP-hard. According to the definition
of the $d$-knapsack constraint, the cardinality constraint with budget $k$
is a special case of a $1$-knapsack constraint when $c(v)=\frac{1}{k},\forall v \in V$.
Because maximizing a submodular function with a cardinality constraint is
NP-hard~\cite{DBLP:journals/mp/NemhauserWF78,DBLP:journals/jacm/Feige98}, \pName is NP-hard as well.
Due to the submodularity of the utility function, a na{\"i}ve approach to \pName
over sliding windows is storing the active window $A_t$ and rerunning a batch algorithm
for \pName on $A_t$ from scratch for every window slide.
Typical batch algorithms for \pName are \algDen~\cite{DBLP:conf/acl/LinB11,DBLP:conf/kdd/LeskovecKGFVG07}
($\shortDen$), an extension of the classic greedy algorithm~\cite{DBLP:journals/mp/NemhauserWF78},
and \textsc{ContinuousGreedy}~\cite{DBLP:conf/soda/KulikST09,DBLP:conf/focs/FeldmanNS11} (\textsc{ContG})
which is based on the multi-linear relaxation technique.
From the theoretical perspective, the approximation ratio of $\shortDen$ for \pName
depends on the dimension of knapsacks $d$ while \textsc{ContG}
can achieve a constant approximation (e.g., $1-\frac{1}{e}-\varepsilon$) independent of $d$.
But \textsc{ContG} suffers from extremely high time
complexity (e.g., $\mathcal{O}(W^{d \cdot \varepsilon^{-4}})$~\cite{DBLP:conf/soda/KulikST09},
see Table~\ref{tbl:comparison}) and is not practical even for very small $W$.
In practice, we implement $\shortDen$ as the batch baseline.
$\shortDen$ returns near-optimal solutions for \pName empirically
when the cost distribution is not extremely
adversary~\cite{DBLP:conf/kdd/LeskovecKGFVG07,DBLP:conf/globalsip/YuXC16}.
Nevertheless, for \pName over sliding windows, $\shortDen$ still needs to scan the
active elements for multiple passes and incurs heavy computational costs.
Hence, our challenge is to design efficient
frameworks to continuously maintain the solutions for \pName over sliding windows
when new elements arrive rapidly,
while guaranteeing a constant approximation ratio w.r.t.~a fixed $d$.

\begin{table}
  \centering
  \caption{Frequently used notations}\label{tbl:notations}
  \begin{tabular}{|c|m{5.35in}|}
    \hline
    \textbf{Notation} & \textbf{Description} \\
    \hline
    $V,v_t$ & $V$ is an unbounded stream of elements; $v_t \in V$ is the $t$-th element in the stream.\\
    \hline
    $d,\xi$ & $d$ is the dimension of the knapsack constraint; $\xi$ is the family of sets defined by the $d$-knapsack constraint.\\
    \hline
    $c_j(v),c_{tj}$ & $c_j(v)$ is the cost of $v$ in the $j$-th knapsack; $c_{tj}$ is the cost of $v_t$ in the $j$-th knapsack.\\
    \hline
    $\gamma_t,\delta_t$ & $\gamma_t=\min_{\forall j \in [d]}c_{tj}$ and $\delta_t=\max_{\forall j \in [d]}c_{tj}$
    are the minimum and maximum costs of $v_t$ in all $d$ knapsacks.\\
    \hline
    $\gamma,\delta$ & $\gamma=\min_{\forall t,\forall j}c_{tj}$ and $\delta=\max_{\forall t,\forall j}c_{tj}$
    are the lower and upper bounds for the costs of any elements in the stream.\\
    \hline
    $f(\cdot),\Delta_f(\cdot|\cdot)$ & $f(\cdot)$ is a monotone submodular utility function;
    $\Delta_f(\cdot|\cdot)$ is the marginal gain defined on $f(\cdot)$.\\
    \hline
    $W$ & $W$ is the size of the sliding window.\\
    \hline
    $A_t$ & $A_t=\langle v_{t'},\ldots,v_t \rangle$ is the active window at time $t$ where $t'=\max\{1,t-W+1\}$.\\
    \hline
    $S^*_t,\optimal_t$ & $S^*_t$ is the optimal solution for \pName w.r.t.~the active window $A_t$ at time $t$;
    $\optimal_t=f(S^*_t)$ denotes the optimal utility value.\\
    \hline
    $S_t$ & $S_t$ denotes an approximate solution for \pName w.r.t.~$A_t$ at time $t$.\\
    \hline
    $X_t,x_i$ & $X_t=\langle x_1,\ldots,x_s \rangle$ is the sequence of $s$ checkpoints at time $t$
    maintained by $\shortWin$ and $\shortWinOpt$; $x_i$ is the $i$-th checkpoint in $X_t$.\\
    \hline
    $S^*_{x,y},S_{x,y}$ & $S^*_{x,y}$ and $S_{x,y}$ are the optimal solution and an approximate solution
    for \pName w.r.t.~a substream $V_{x,y}=\langle v_{x},\ldots,v_{y} \rangle$.\\
    \hline
  \end{tabular}
\end{table}

Before moving on to the subsequent sections,
we summarize the frequently used notations in Table~\ref{tbl:notations}.

\section{Applications}\label{sec:application}

In this section, we give two examples of \rss applications and
describe how they are modeled as \pName over sliding windows.
The experiments for both applications in real-world datasets will be reported in Section~\ref{sec:experiment}.
Note that many more \rss problems
can also be modeled as \pName (see Section~\ref{sec:related}),
which could potentially benefit from this work.

\subsection{Social Stream Summarization}\label{subsec:social:stream}

Massive data is continuously generated as a stream by hundreds of
millions of users on social platforms, e.g., Twitter.
\emph{Social stream summarization} aims to retain a small portion of representative elements
from a user-generated stream. One common approach is topic-preserving
summarization~\cite{DBLP:conf/cikm/ZhuangRHGHA16,DBLP:conf/ijcai/TamWTYH17}
that selects a subset of posts that best preserve \emph{latent topics} in the stream.
We focus on topic-preserving summarization in the sliding window model
to capture the evolving nature of social streams,
i.e., topics under discussion change over time~\cite{DBLP:conf/ijcai/TamWTYH17}.
We consider a collection of social posts $V$ is available as a stream in ascending order of
timestamp. A social post $v \in V$ is represented as a bag of $l$ words $\{w_1,\ldots,w_l\}$
drawing from the vocabulary $\mathcal{W}$.
The utility $f(S)$ for a set of elements $S$ is computed by
summing up the weights of words in $S$
where the weight of a word $w$ is acquired based on its information entropy~\cite{DBLP:conf/cikm/ZhuangRHGHA16}.
Specifically,
\begin{equation}\label{eq:social}
  f(S) = \sum_{w \in \mathcal{W}} \max_{v \in S} n(v,w) \cdot p(w) \cdot \log\frac{1}{p(w)}
\end{equation}
where $n(v,w)$ is the frequency of word $w$ in element $v$,
$p(w)=\frac{\sum_{v \in V}n(v,w)}{\sum_{v \in V}\sum_{w \in \mathcal{W}}n(v,w)}$
is the probability of generating a word $w$ from the topic model.
$f(S)$ has been proved to be monotone and submodular~\cite{DBLP:conf/cikm/ZhuangRHGHA16}.
Furthermore, the representatives should satisfy the following $3$-knapsack constraint.
First, a uniform cost $c_1(v)$ is assigned to each element $v \in V$,
i.e., $c_1(v)=\frac{1}{k}$, to bound the size of the representative set within $k$~\cite{DBLP:conf/cikm/ZhuangRHGHA16,DBLP:conf/ijcai/TamWTYH17}.
Second, a cost $c_2(v)$ is assigned to the length $l$ of element $v$
since users prefer shorter summaries to longer ones~\cite{DBLP:conf/acl/LinB11,DBLP:conf/naacl/LinB10}.
For normalization, we compute the average number of words $\overline{l}$ in one element
and assign $c_2(v)$ as follows: given an element $v$ of $l$ words, $c_2(v)=\frac{1}{k}\cdot\frac{l}{\overline{l}}$.
For example, when $\overline{l}=5, k=10$, an element $v$ with $l=10$ words
has a cost $c_2(v)=0.2$. Third, a cost $c_3(v)$ is assigned according to social influence~\cite{DBLP:conf/cikm/ZhuangRHGHA16,DBLP:conf/globalsip/YuXC16}.
Let $fl(v)$ denote the number of followers of the user who posts $v$.
We consider $c_3(v) = \min(\delta, \frac{1}{k} \cdot \frac{\log (1 + fl(v))}{\log (1+\overline{fl})})$
where $\overline{fl}$ is the average number of followers of each user
and $\delta$ is the upper-bound cost.
We assign lower costs to the elements posted by more influential users
so that the summary could include more influential elements.
The upper-bound cost $\delta$ is assigned to elements posted by users with very few (e.g., 0 or 1)
followers for normalization.
To sum up, the social stream summarization is modeled as maximizing $f(\cdot)$
in Equation~\ref{eq:social} with a $3$-knapsack constraint defined by $c_1(\cdot)$, $c_2(\cdot)$, and $c_3(\cdot)$
over the active window $A_t$.

\subsection{Active Set Selection}
\label{subsec:active:set}

\emph{Active set selection}~\cite{DBLP:conf/kdd/BadanidiyuruMKK14,DBLP:conf/icml/GomesK10}
is a common approach to scaling kernel methods to massive datasets.
It aims to select a small subset of elements with the maximal information entropy
from the source dataset.
In some sites like \emph{Yahoo!}, weblogs are continuously generated by users as a stream.
Given a stream of weblogs $V$,
each record $v \in V$ is modeled as a multi-dimensional feature vector.
The representativeness of a set of vectors $S$
is measured by the Informative Vector Machine~\cite{DBLP:conf/nips/LawrenceSH02} (IVM):
\begin{equation}\label{eq:IVM}
  f(S)=\frac{1}{2}\log\det(\mathbf{I}+\sigma^{-2}\mathbf{K}_{S,S})
\end{equation}
where $\mathbf{K}_{S,S}$ is an $|S|\times|S|$ kernel matrix indexed by $S$
and $\sigma>0$ is a regularization parameter.
For each pair of elements $v_i,v_j \in S$, the $(i,j)$-th entry $\mathcal{K}_{i,j}$ of $\mathbf{K}$
represents the similarity between $v_i$ and $v_j$ measured via
a symmetric positive definite kernel function.
We adopt the squared exponential kernel embedded in the Euclidean space,
i.e., $\mathcal{K}_{i,j}=\exp(-\frac{\lVert v_i-v_j \rVert^2_2}{h^2})$.
It has been proved that $f(\cdot)$ in Equation~\ref{eq:IVM}
is a monotone submodular function~\cite{DBLP:conf/icml/GomesK10}.
Furthermore, other than assigning a fixed cost to each feature vector,
existing methods also use different schemes to assign costs,
e.g., generating from a Gamma distribution or marginal-dependent costs~\cite{DBLP:conf/nips/CuongX16}.
Thus, we consider a more general case:
each feature vector $v$ is associated with a cost $c(v)$
drawing from an arbitrary distribution $\mathcal{D}$ within range $(0,1)$.
The objective is to select a subset $S$ of feature vectors
such that $f(S)$ in Equation~\ref{eq:IVM} is maximized
subject to a $1$-knapsack constraint defined by $c(\cdot)$ over the active window $A_t$.

\section{The KnapWindow Framework}\label{sec:framework}

\begin{figure}
  \centering
  \includegraphics[width=0.6\textwidth]{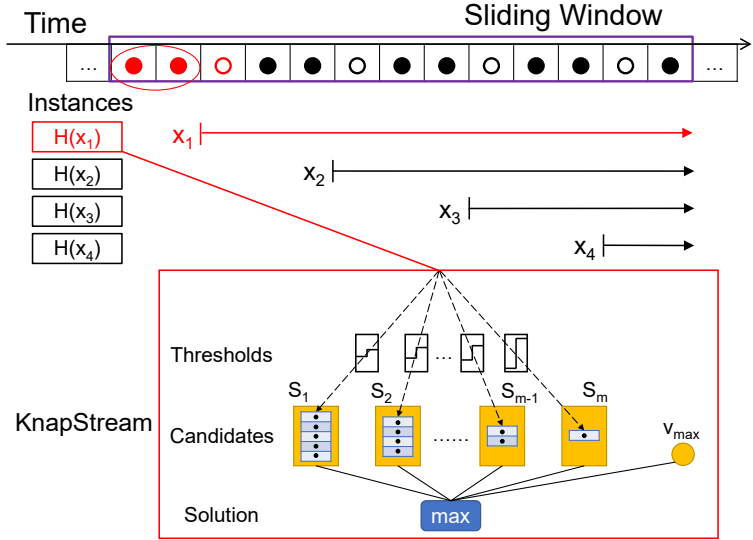}
  \caption{An illustration of the \algWindow framework.}
  \label{fig:kw-framework}
\end{figure}

In this section, we propose the \algWindow ($\shortWin$) framework for \pName
over sliding windows. The architecture of $\shortWin$ is
illustrated in Figure~\ref{fig:kw-framework}.
$\shortWin$ always stores all active elements in $A_t$ at any time $t$.
Then, $\shortWin$ adapts the \algStream ($\shortKs$) algorithm that
provides an approximation solution for \pName in append-only streams
to work in the sliding model in the following manner.
It maintains a sequence of \emph{checkpoints}
$X_t=\langle x_1,\ldots,x_s \rangle \subseteq [t',t]$ over the active window $A_t$.
The interval between any neighboring checkpoints $x_i$ and $x_{i+1}$ is equal
(e.g., the interval is $3$ in Figure~\ref{fig:kw-framework}).
For each checkpoint $x_i$, a $\shortKs$ instance $\mathcal{H}(x_i)$
is maintained by processing an append-only stream from $v_{x_i}$ to $v_t$.
To retrieve the solution for \pName at time $t$, $\shortWin$ always uses the result from
$\mathcal{H}(x_1)$ corresponding to $x_1$.
$\mathcal{H}(x_1)$ first post-processes the active elements before
$v_{x_1}$ (e.g., the solid red ones in Figure~\ref{fig:kw-framework})
and uses the result after post-processing as the final solution.

The scheme of $\shortKs$ to maintain a solution for \pName over an append-only stream
is also illustrated in Figure~\ref{fig:kw-framework}.
First, $\shortKs$ approximates the optimal utility $\optimal$ for \pName by a sequence of estimations.
Then, $\shortKs$ maintains a \emph{candidate} for each estimation with a unique \emph{threshold}
derived from the estimation.
Whenever receiving a new element, $\shortKs$ checks whether it can be included
into each candidate independently according to the threshold.
Finally, $\shortKs$ selects the candidate with the maximum utility
among all candidates as the solution for its processed substream.

Next, Section~\ref{subsec:ks} will present the \algStream algorithm for \pName in append-only
streams. Then, Section~\ref{subsec:kw} will introduce how the \algWindow
algorithm adapts \algStream for the sliding window model.
Finally, Section~\ref{subsec:kw:theoretical} will analyze both algorithms theoretically.

\subsection{The KnapStream Algorithm}\label{subsec:ks}

In this subsection, we propose the \algStream ($\shortKs$) algorithm to maintain a solution for \pName
w.r.t.~an append-only stream $V_{x,y}=\langle v_x,\ldots,v_y \rangle$ from time $x$ to $y$.
$\shortKs$ follows the threshold-based
framework~\cite{DBLP:conf/kdd/BadanidiyuruMKK14,DBLP:journals/topc/KumarMVV15}
for streaming submodular maximization. Its mechanism depends
on estimating the optimal utility value $\optimal$ for \pName w.r.t.~$V_{x,y}$.
Although $\optimal$ cannot be exactly determined unless P=NP,
$\shortKs$ tracks the lower and upper bounds for $\optimal$ from the observed elements online
and maintains a sequence of candidates with different estimations
for $\optimal$ in the range.
Each candidate derives a unique threshold for the marginal gain
according to its estimation for $\optimal$.
When a new element arrives, a candidate decides whether to include it
based on the marginal gain of adding it into the candidate and the candidate's threshold.
After processing the stream, the candidate with the maximum utility
is used as the solution.

Although having a similar scheme,
the algorithms in~\cite{DBLP:conf/kdd/BadanidiyuruMKK14} and~\cite{DBLP:journals/topc/KumarMVV15}
only work with one cardinality constraint,
whereas $\shortKs$ is different from them in two aspects
to achieve an approximation guarantee for general $d$-knapsack constraints:
(1) the criterion for the inclusion of an element considers not only its marginal gain
but also its costs, i.e., it checks the cost-effectiveness of adding the element
in each knapsack and includes it only when its cost-effectiveness
reaches the threshold in $d$ knapsacks;
(2) the singleton element with the maximum self-utility is also a candidate solution.

\begin{algorithm}
\begin{algorithmic}[1]
  \Require A stream $V_{x,y}=\langle v_x,\ldots,v_y \rangle$, a parameter $\lambda$
  \Ensure The solution $S_{x,y}$ for \pName w.r.t.~$V_{x,y}$
  \State $\Phi=\langle (1+\lambda)^l|l \in \mathbb{Z} \rangle$
  \State \textbf{for all} $\phi \in \Phi$ \textbf{do} $S_{\phi} \gets \emptyset$
  \State Initialize $m,M \gets 0$ and $v_{max} \gets nil$
  \For{$t \gets x,\ldots,y$}
    \State \textbf{if} $f(\{v_t\})>f(\{v_{max}\})$ \textbf{then} $v_{max} \gets v_t$ \label{line:ks:variables:start}
    \State $\delta_t=\max_{\forall j \in [d]}c_{tj}, \gamma_t=\min_{\forall j \in [d]}c_{tj}$
    \label{line:ks:stream:start}
    \If{$\frac{f(\{v_t\})}{\gamma_t}>M$}
      \State $M \gets \frac{f(\{v_t\})}{\gamma_t}, m \gets f(\{v_t\})$ \label{line:ks:mM}
    \EndIf
    \label{line:ks:variables:end}
    \State $\Phi_t= \langle (1+\lambda)^l|l\in\mathbb{Z}, m\leq(1+\lambda)^l\leq M \cdot (1+d) \rangle$
    \label{line:ks:candidates:start}
    \State Delete $S_{\phi}$ if $\phi \notin \Phi_t$
    \label{line:ks:candidates:end}
    \ForAll{$\phi \in \Phi_t$}
    \label{line:ks:element:start}
      \If{$\Delta_f(v_t|S_{\phi}) \geq \frac{\delta_t \cdot \phi}{1+d} \wedge S_{\phi}\cup\{v_t\} \in \xi$
      \label{line:ks:threshold}}
        \State $S_{\phi} \gets S_{\phi} \cup \{v_t\}$
      \EndIf
    \EndFor
    \label{line:ks:stream:end}
    \label{line:ks:element:end}
  \EndFor
  \State $S_{max} \gets \argmax_{\phi \in \Phi}f(S_{\phi})$
  \label{line:ks:streamquery:start}
  \State \Return{$S_{x,y} \gets \argmax (f(S_{max}),f(\{v_{max}\}))$}
  \label{line:ks:streamquery:end}
\end{algorithmic}
\caption{\algStream}\label{alg:ks}
\end{algorithm}

The pseudo-code of $\shortKs$ is presented in Algorithm~\ref{alg:ks}.
Three auxiliary variables are maintained by $\shortKs$
(Lines~\ref{line:ks:variables:start}--\ref{line:ks:variables:end}):
$v_{max}$ stores the element with the maximum self-utility;
$M$ and $m$ track the upper and lower bounds for $\optimal$.
Specifically, $M$ is the maximum cost-effectiveness any observed element can achieve
and $m$ is the corresponding self-utility.
We will explain why they are the upper and lower bounds for $\optimal$
in the proof of Theorem~\ref{thm:ksApprox}.
The sequence of estimations $\Phi=\langle (1+\lambda)^l|l\in\mathbb{Z}, m\leq(1+\lambda)^l\leq M \cdot (1+d) \rangle$
and corresponding candidates are updated based on the up-to-date $m$ and $M$
(Lines~\ref{line:ks:candidates:start}--\ref{line:ks:candidates:end}).
Then, given an element $v_t$, each candidate checks whether to include it independently.
For each $\phi \in \Phi_t$,
if the marginal gain $\Delta_f(v_t|S_{\phi})$ of adding $v_t$ to $S_{\phi}$
reaches $\frac{\delta_t \cdot \phi}{1+d}$ where $\delta_t=\max_{\forall j \in [d]}c_{tj}$
and the $d$-knapsack constraint is still satisfied after adding $v_t$,
$v_t$ will be included into $S_{\phi}$
(Lines~\ref{line:ks:element:start}--\ref{line:ks:element:end}).
Finally, after processing every element in the stream,
it first finds $S_{max}$ with the maximum utility among the candidates
and then compares the utility of $S_{max}$ with that of $\{v_{max}\}$.
The one with the higher utility is returned as
the solution $S_{x,y}$ for \pName w.r.t.~the stream $V_{x,y}$
(Lines~\ref{line:ks:streamquery:start} and~\ref{line:ks:streamquery:end}).

\subsection{The KnapWindow Algorithm}\label{subsec:kw}

In this subsection, we present the \algWindow ($\shortWin$) algorithm.
It adapts $\shortKs$ for \pName in the sliding window model
by maintaining a sequence of \emph{checkpoints} and corresponding
$\shortKs$ instances over the sliding window.
At any time $t$, $\shortWin$ maintains a sequence of $s$ checkpoints
$X_t=\langle x_1,\ldots,x_s \rangle \subseteq [t',t]$.
The interval between any neighboring checkpoints in $X_t$ is always equal.
Given the interval $L \in \mathbb{Z}^{+}$, $\shortWin$ only creates a new checkpoint and
initiates a new $\shortKs$ instance for every $L$ elements.
For each checkpoint $x_i$, a $\shortKs$ instance $\mathcal{H}(x_i)$ is maintained by
processing a substream from element $v_{x_i}$ to the up-to-date element $v_t$.
Whenever the first checkpoint $x_1$ expires from the sliding window ($x_1 < t'$ where $t' = max(1,t-W+1)$),
it will be deleted from $X_t$.
The corresponding $\shortKs$ instance $\mathcal{H}(x_1)$ will be terminated as well.
To provide the solution for \pName w.r.t.~$A_t$, it uses the result from $\mathcal{H}(x_1)$.
But it is noted that the elements from $v_{t'}$ to $v_{x_{1}-1}$ have not been processed
by $\mathcal{H}(x_1)$ yet. Therefore, it feeds the unprocessed elements to $\mathcal{H}(x_1)$
before returning the final solution.

\begin{algorithm}
\begin{algorithmic}[1]
  \Require A stream $V=\langle v_1,v_2,\ldots\rangle$, the window size $W$, the interval $L$ for neighboring checkpoints
  \Ensure The solution $S_t$ for \pName at time $t$
  \State Initialize $s \gets 0$, $X_0 \gets \emptyset$
  \For{$t \gets 1,2,\ldots$}
    \If{$t \in \{x | x=j \cdot L, j \in \mathbb{N}\}$}
    \label{line:kw:create:start}
        \State $s \gets s+1$, $x_s \gets t$, and $X_t \gets X_{t-L} \circ \langle x_s \rangle$
        \State Initiate a $\shortKs$ instance $\mathcal{H}(x_s)$
    \EndIf
    \label{line:kw:create:end}
    \While{$t > W \wedge x_1 < t'$}
    \label{line:kw:expire:start}
        \State $X_t \gets X_t \setminus \langle x_1 \rangle$, terminate $\mathcal{H}(x_1)$
        \State Shift the remaining checkpoints, $s \gets s-1$
    \EndWhile
    \label{line:kw:expire:end}
    \For{$i \gets 1,\ldots,s$}
    \label{line:kw:process:start}
        \State $\mathcal{H}(x_i)$ processes $v_t$ according to Algorithm~\ref{alg:ks}
    \EndFor
    \label{line:kw:process:end}
    \State \textit{// The post-processing procedure at time $t$}
    \State $\mathcal{H}(x_1)$ processes each element from $v_{t'}$ to $v_{x_{1}-1}$ according to Algorithm~\ref{alg:ks}
    \label{line:kw:post}
    \State \Return $S_t \gets$ the solution of $\mathcal{H}(x_1)$
    \label{line:kw:result}
  \EndFor
\end{algorithmic}
\caption{\algWindow}\label{alg:kw}
\end{algorithm}

The pseudo-code of $\shortWin$ is presented in Algorithm~\ref{alg:kw}.
The sequence of checkpoints is initialized to $X_0=\varnothing$.
A checkpoint $x_s=t$ is created and appended to the end of $X_t$ at time $t=L,2L,\ldots$.
A $\shortKs$ instance $\mathcal{H}(x_s)$ is initiated accordingly
(Lines~\ref{line:kw:create:start}--\ref{line:kw:create:end}).
Then, it deletes the expired checkpoints
from $X_t$ (Lines~\ref{line:kw:expire:start}--\ref{line:kw:expire:end}).
Subsequently, each checkpoint processes $v_t$
and updates the result independently.
This procedure follows Lines~\ref{line:ks:stream:start}--\ref{line:ks:stream:end}
of Algorithm~\ref{alg:ks}.
To provide the solution $S_t$ for \pName at time $t$,
$\mathcal{H}(x_1)$ post-processes the elements from $v_{t'}$ to $v_{x_{1}-1}$
(Line~\ref{line:kw:post}).
Finally, the solution of $\mathcal{H}(x_1)$ after post-processing
is returned as $S_t$ (Line~\ref{line:kw:result}).

\subsection{Theoretical Analysis}\label{subsec:kw:theoretical}

In this subsection, we analyze the approximation ratios and complexities
of $\shortKs$ and $\shortWin$. In the theoretical analysis, we assume
the cost of any element is bounded by $\gamma$ and $\delta$,
i.e., $0<\gamma\leq c_{tj}\leq\delta\leq1$ for all $t,j$.
It is noted that the algorithms do not need to know $\gamma$ and $\delta$
in advance.

The roadmap of our analysis is as follows.
First of all, we present the approximation ratio of $\shortKs$.
We first show that if we knew the optimal utility $\optimal$ for \pName w.r.t.~$V_{x,y}$
in advance, the candidate whose estimation is the closest to $\optimal$
would be a $\frac{(1-\lambda)(1-\delta)}{1+d}$ approximate solution
(Lemma~\ref{lm:knownOPT}). However, the approximation ratio depends
on $\delta$ and may degrade arbitrarily when $\delta$ increases.
Therefore, we further show that if the singleton element with the
maximum self-utility is also considered as a candidate solution
(Line~\ref{line:ks:streamquery:start} of Algorithm~\ref{alg:ks}),
there is a lower bound for the approximation ratio
regardless of $\delta$ (Lemma~\ref{lm:gamma}).
Then, as $\optimal$ is unknown unless P=NP, we analyze how $\shortKs$
can track the lower and upper bounds for $\optimal$ and how many
different estimations are required to guarantee that at least one
of them approximates $\optimal$ within a bounded error ratio
(Theorem~\ref{thm:ksApprox}). As $\shortKs$ maintains one candidate
for each $\optimal$ estimation, we can get its time and space complexity accordingly.
After providing the theoretical results for $\shortKs$,
we extend these results to $\shortWin$.
Specifically, $\shortWin$ retains the approximation ratio of $\shortKs$
because it is guaranteed that the solution of $\shortWin$ is returned
only after processing all active elements (Theorem~\ref{thm:kwApprox}).
Finally, we analyze the complexity of $\shortWin$.

\begin{lemma}
\label{lm:knownOPT}
  Assuming there exists $\phi \in \Phi$ such that $(1-\lambda) \optimal \leq \phi \leq \optimal$
  where $\optimal$ is the optimal utility of \pName w.r.t.~$V_{x,y}$,
  $S_{\phi}$ satisfies that $f(S_{\phi}) \geq \frac{(1-\lambda)(1-\delta)}{1+d} \cdot \optimal$.
\end{lemma}
\begin{proof}
  Let $s_i$ be the $i$-th element added to $S_{\phi}$,
  $S^i_{\phi}$ be $\{s_1,\ldots,s_i\}$ for $i\in[0,|S_{\phi}|]$ with
  $S^0_{\phi}=\emptyset$, $b_j=c_j(S_{\phi})$ for $j \in [d]$ be the
  cost of $S_{\phi}$ in the $j$-th knapsack,
  and $b=\max_{j\in[d]}b_j$ be the maximal cost of $S_{\phi}$ among $d$ knapsacks.
  According to Line~\ref{line:ks:threshold} in Algorithm~\ref{alg:ks},
  we have $\Delta_f(s_i|S^{i-1}_{\phi}) \geq \frac{c_j(s_i) \cdot \phi}{1+d}$
  for $j\in[d]$. It holds that:
  \begin{displaymath}
    f(S_{\phi}) = \sum^{|S_{\phi}|}_{i=1} \Delta_f(s_i|S^{i-1}_{\phi})
    \geq \frac{\phi}{1+d} \cdot c_j(S_{\phi}) = \frac{\phi}{1+d} \cdot b_j
  \end{displaymath}
  Therefore, $f(S_{\phi}) \geq \frac{\phi}{1+d} \cdot b$.

  Next, we discuss two cases separately as follows.

  \textbf{Case 1}. When $b\geq(1-\delta)$, we have:
  \begin{displaymath}
    f(S_{\phi}) \geq \frac{b\cdot\phi}{1+d} \geq \frac{(1-\delta)\cdot\phi}{1+d}
    \geq \frac{(1-\lambda)(1-\delta)}{1+d}\cdot\optimal
  \end{displaymath}

  \textbf{Case 2}. When $b<(1-\delta)$, we have $\forall v \in V \setminus S_{\phi}$,
  $S_{\phi}\cup\{v\}\in\xi$. Let $S^*$ be the optimal solution for $V$
  and $a$ be an element in $S^* \setminus S_{\phi}$.
  Since $a$ is not added to $S_{\phi}$, there must exist $\mu(a)\in[d]$ such that
  $\Delta_f(a|S'_{\phi}) < \frac{c_{\mu(a)}(a)\cdot\phi}{1+d}$,
  where $S'_{\phi} \subseteq S_{\phi}$ is the subset of $S_{\phi}$ when $a$ is processed.
  We consider $S^*_j=\{a|a \in S^* \setminus S_{\phi}
  \wedge \mu(a)=j\}$ for $j\in[d]$.
  Due to the submodularity of $f(\cdot)$, we acquire:
  \begin{displaymath}
    f(S_{\phi} \cup S^*_j) - f(S_{\phi})
    \leq \sum_{a \in S^*_j} \Delta_f(a|S_{\phi})
    < \frac{\phi \cdot c_j(S^*_j)}{1+d} \leq \frac{\phi}{1+d}
  \end{displaymath}
  Then, because $S^* \setminus S_{\phi}=\cup^d_{j=1}S^*_j$, we have:
  \begin{displaymath}
    f(S^* \cup S_{\phi})-f(S_{\phi}) \leq
    \sum^d_{j=1} f(S_{\phi} \cup S^*_j)-f(S_{\phi}) < \frac{d\phi}{1+d}
  \end{displaymath}
  Finally, we get $f(S_{\phi}) > \optimal-\frac{d}{1+d}\optimal \geq \frac{1}{1+d}\cdot\optimal$.

  Considering both cases, we conclude the proof.
\end{proof}

Lemma~\ref{lm:knownOPT} has proved that $\shortKs$ achieves a good approximation ratio
when $\delta$ is small.
Next, we further analyze the case where $\delta>0.5$
and prove that the approximation ratio has a lower bound
regardless of $\delta$.

\begin{lemma}\label{lm:gamma}
  When $\delta>0.5$, it satisfies that at least one of $f(S_{\phi})$ and $f(\{v_{max}\})$
  is greater than $\frac{0.5(1-\lambda)}{1+d}\cdot\optimal$.
\end{lemma}
\begin{proof}
  Lemma~\ref{lm:gamma} naturally follows when $b \geq 0.5$
  (Case 1 of Lemma~\ref{lm:knownOPT})
  or for all $a \in S^* \setminus S_{\phi}$, $a$ is excluded from $S_{\phi}$
  because its marginal gain does not reach the threshold in some knapsack
  (Case 2 of Lemma~\ref{lm:knownOPT}).

  Thus, we only need to consider the following case:
  there exists some elements whose marginal gains reach the threshold in all knapsacks
  but are excluded from $S_{\phi}$ because including them into $S_{\phi}$ violates
  the $d$-knapsack constraint.
  Assuming $a$ is such an element for $S_{\phi}$,
  we have $\Delta_f(a|S'_{\phi}) \geq \frac{c_j(a)\cdot\phi}{1+d}$
  and $c_j(S'_{\phi})+c_j(a)>1$ for some $j\in[d]$.
  In this case, we have:
  \begin{displaymath}
    f(S'_{\phi}\cup\{a\}) \geq \frac{\phi}{1+d} \cdot \Big( c_j(S'_{\phi})+c_j(a) \Big) > \frac{\phi}{1+d}
  \end{displaymath}
  Due to the monotonicity and submodularity of $f(\cdot)$, we get:
  \begin{displaymath}
    \frac{\phi}{1+d} \leq f(S'_{\phi}\cup\{a\}) \leq f(S'_{\phi}) + f(\{a\}) \leq f(S_{\phi}) + f(\{v_{max}\})
  \end{displaymath}
  Therefore, at least one of $f(S_{\phi})$ and $f(\{v_{max}\})$
  is greater than $\frac{0.5\phi}{1+d}\cdot\optimal$ and we conclude the proof.
\end{proof}

Given Lemmas~\ref{lm:knownOPT} and~\ref{lm:gamma}, we prove that $\shortKs$ achieves an
approximation factor of $\frac{(1-\delta)(1-\lambda)}{1+d}$ (when $\delta \leq 0.5$) or
$\frac{0.5(1-\lambda)}{1+d}$ (when $\delta>0.5$).

\begin{theorem}\label{thm:ksApprox}
  The solution $S_{x,y}$ returned by Algorithm~\ref{alg:ks} satisfies
  $f(S_{x,y}) \geq \frac{1-\varepsilon}{1+d} \cdot f(S^*_{x,y})$
  where $S^*_{x,y}$ is the optimal solution for \pName w.r.t.~$V_{x,y}$
  and $\varepsilon = \min(\delta+\lambda,0.5+\lambda)$.
\end{theorem}
\begin{proof}
  By Lemmas~\ref{lm:knownOPT} and~\ref{lm:gamma},
  we can say Theorem~\ref{thm:ksApprox} naturally holds
  if there exists at least one $\phi \in \Phi$
  such that $(1-\lambda)\optimal\leq\phi\leq\optimal$.
  First, we show $m$ and $M$ are the lower and upper bounds for $\optimal$.
  It is easy to see $m \leq \optimal$ as $m \leq f(\{v_{max}\})$ and $\{v\}\in\xi$
  for any $v \in V$.
  $M$ maintains the maximum cost-effectiveness among all elements.
  We have $M \geq \frac{f(\{v_i\})}{c_{ij}}$, $\forall i\in[x,y]$ and $\forall j\in[d]$.
  Let $S^*_{x,y}=\{a_1,\ldots,a_{|S^*|}\}$ be the optimal solution for $V_{x,y}$.
  As $f(\cdot)$ is monotone submodular,
  $\optimal \leq \sum^{|S^*_{x,y}|}_{i=1}f(\{a_i\}) \leq c_j(S^*) M$ for $j\in[d]$.
  As $c_j(S^*_{x,y}) \leq 1$, we have $M \geq \optimal$.
  $\shortKs$ estimates $\optimal$ by a sequence
  $\langle (1+\lambda)^l|l\in\mathbb{Z},m \leq (1+\lambda)^l \leq M(1+d) \rangle$.
  Then, there exists at least one estimation $\phi$ such that $\phi\leq\optimal\leq(1+\lambda)\phi$.
  Equivalently, $(1-\lambda)\optimal\leq\phi\leq\optimal$.
  Therefore, we conclude the proof by combining this result
  with Lemma~\ref{lm:knownOPT} and~\ref{lm:gamma}.
\end{proof}

\textbf{The Complexity of $\shortKs$.}
As only one pass over the stream is permitted,
to avoid missing elements with marginal gains of greater than $\frac{M}{1+d}$,
$\shortKs$ maintains the candidates for estimations within an increased
range $[m,(1+d)M]$ instead of $[m,M]$.
Then, because $\frac{m}{M}\leq\gamma$ (Line~\ref{line:ks:mM} of Algorithm~\ref{alg:ks}), the number of candidates in $\shortKs$
is bounded by $\lceil \log_{1+\lambda}\gamma^{-1}(1+d) \rceil$.
Thus, we have $\shortKs$ maintains $\mathcal{O}(\frac{\log(d\cdot\gamma^{-1})}{\varepsilon})$ candidates.
For each candidate, one function call is required to evaluate whether to add a new element.
Thus, the time complexity to update one element is $\mathcal{O}(\frac{\log(d\cdot\gamma^{-1})}{\varepsilon})$.
Finally, at most $\gamma^{-1}$ elements can be maintained in each candidate.
Otherwise, the $d$-knapsack constraint must not be satisfied.
Therefore, the number of elements stored is $\mathcal{O}(\frac{\log(d\cdot\gamma^{-1})}{\gamma\cdot\varepsilon})$.

Next, we present the approximation factor of $\shortWin$.
\begin{theorem}\label{thm:kwApprox}
  The solution $S_{t}$ returned by Algorithm~\ref{alg:kw} satisfies
  $f(S_{t}) \geq \frac{1-\varepsilon}{1+d} \cdot \optimal_t$
  where $\optimal_t$ is the optimal utility for \pName w.r.t.~$A_t$ at time $t$
  and $\varepsilon = \min(\delta+\lambda,0.5+\lambda)$.
\end{theorem}
It is obvious that $\mathcal{H}(x_1)$ must have
processed every element in $A_t$ after post-processing.
As the approximation ratio of $\shortKs$ is order-independent,
i.e., no assumption is made for the arrival order of elements,
Theorem~\ref{thm:kwApprox} holds.

\textbf{The Complexity of $\shortWin$.}
$\shortWin$ maintains $s=\lceil \frac{W}{L} \rceil$ checkpoints for $A_t$
and thus updates the $\shortKs$ instances for an element
in $\mathcal{O}(\frac{s \cdot \log(d \cdot \gamma^{-1})}{\lambda})$ time.
In addition, it takes $\mathcal{O}(\frac{L \cdot \log(d \cdot \gamma^{-1})}{\lambda})$
time for post-processing.
The time complexity of $\shortWin$ is $\mathcal{O}(\frac{(s+L) \cdot \log(d \cdot \gamma^{-1})}{\lambda})$.
When $s=L=\sqrt{W}$, it becomes $\mathcal{O}(\frac{\sqrt{W} \cdot \log(d \cdot \gamma^{-1})}{\lambda})$.
Finally, because all active elements must be stored,
the space complexity of $\shortWin$ is $\mathcal{O}(W)$.

\section{The KnapWindowPlus Framework}\label{sec:kwplus}

Although $\shortWin$ can provide approximation solutions for \pName with a theoretical bound,
it still suffers from two drawbacks that
limit its application for a large window size $W$.
First, $\shortWin$ cannot handle the case when the window does not fit in the main memory.
Second, as $\mathcal{O}(\sqrt{W})$ $\shortKs$ instances are maintained for a size-$W$ sliding window,
the efficiency of $\shortWin$ degrades with increasing $W$.
To improve upon $\shortWin$, we further propose
the \algWindowOpt framework ($\shortWinOpt$) in this section.

\begin{figure}
  \centering
  \includegraphics[width=0.6\textwidth]{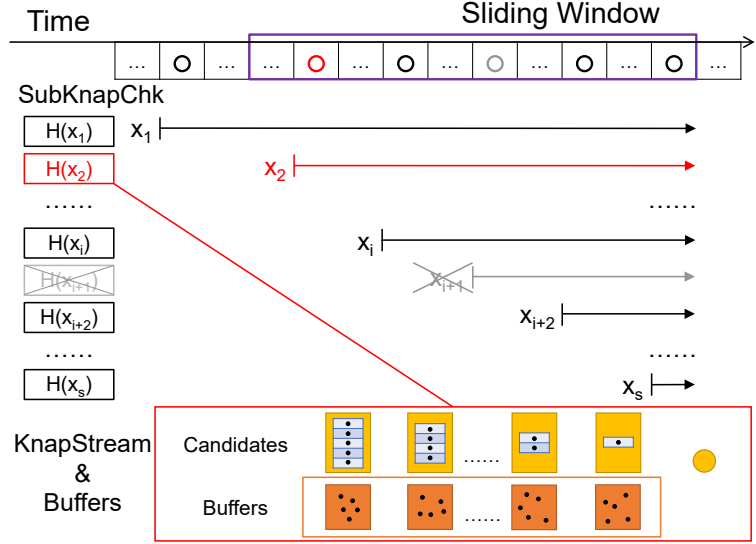}
  \caption{An illustration of the \algWindowOpt framework}
  \label{fig:kwplus-framework}
\end{figure}

The architecture of $\shortWinOpt$ is illustrated in Figure~\ref{fig:kwplus-framework}.
The basic idea of $\shortWinOpt$ is similar to $\shortWin$:
it also keeps the sequence of checkpoints $X_t=\langle x_1,\ldots,x_s \rangle$
and maintains a $\shortKs$ instance to process a substream
from $v_{x_i}$ to $v_t$ at time $t$ in each checkpoint $x_i \in X_t$.
However, $\shortWinOpt$ is substantially different from $\shortWin$ in the following four aspects.
First, $\shortWinOpt$ does not store the entire active window
but only keeps the elements within each $\shortKs$ instance.
The number of elements kept by $\shortWinOpt$ is
empirically much smaller than $W$.
Second, $\shortWinOpt$ builds an index \histogram for checkpoint maintenance.
Instead of maintaining a sequence of checkpoints with equal interval,
$\shortWinOpt$ creates a checkpoint and the corresponding $\shortKs$ instance for every arrival element.
Then, \histogram manages the checkpoints based on their utilities and deletes
a checkpoint whenever it can be approximated by its successors.
By using \histogram, the number of checkpoints in $\shortWinOpt$ is independent of $W$.
Third, $\shortWinOpt$ will keep one expired checkpoint (i.e., $x_1$) when $t>W$.
It tracks the optimal utility $\optimal_{t}$ for \pName w.r.t.~$A_t$
to guarantee the theoretical soundness of the solutions.
Fourth, $\shortWinOpt$ maintains a buffer with tunable size along with each candidate
of the $\shortKs$ instances. In the post-processing procedure, the elements in buffers are
added into the candidates to improve the utilities of solutions.

Next, Section~\ref{subsec:kwplus} will introduce the $\shortWinOpt$ algorithm.
Then, Section~\ref{subsec:kwplus:analysis} will provide a theoretical analysis for $\shortWinOpt$.
Finally, Section~\ref{subsec:discussion} will discuss how to adapt $\shortWin$ and $\shortWinOpt$
to the scenario where the sliding window shifts for more than one element at a time.

\subsection{The KnapWindowPlus Algorithm}\label{subsec:kwplus}

In this subsection, we describe the \algWindowOpt algorithm ($\shortWinOpt$) in detail.
We first present a novel index \emph{Submodular Knapsack Checkpoints} (\histogram)
to maintain a sequence of \emph{checkpoints} and corresponding $\shortKs$ instances
over the sliding window.
Then, we show the procedures for buffer maintenance and post-processing.

\textbf{Submodular Knapsack Checkpoints.}
At time $t$, an index called \emph{Submodular Knapsack Checkpoints} (\histogram)
comprises a sequence of $s$ checkpoints $X_t=\langle x_1,\ldots,x_s \rangle$ where $x_1<\ldots<x_s=t$.
For each checkpoint $x_i$, a $\shortKs$ instance $\mathcal{H}(x_i)$ is maintained.
$\mathcal{H}(x_i)$ processes a substream from $v_{x_i}$ to $v_t$
and will be terminated when $x_i$ is deleted from \histogram.
When $t>W$, the first checkpoint $x_1$ expires (i.e., $x_1<t'$)
but is not deleted from \histogram immediately.
It is maintained to track the upper bound for
the optimal utility $\optimal_t$ of \pName w.r.t.~$A_t$.
However, the result of $\mathcal{H}(x_1)$
cannot be used as the solution at time $t$ in this case
because it may contain expired elements.
\histogram restricts the number of expired checkpoints to at most $1$.
Therefore, $x_2$ must not expire and the result of $\mathcal{H}(x_2)$
is returned as the solution for \pName w.r.t.~$A_t$ when $t>W$.

The idea of maintaining a sequence of \emph{checkpoints} over sliding windows
is inspired by \emph{smooth histograms}~\cite{DBLP:conf/focs/BravermanO07}.
However, according to the analysis in~\cite{DBLP:conf/www/EpastoLVZ17},
the method in~\cite{DBLP:conf/focs/BravermanO07} cannot be directly applied to \pName
because it requires an append-only streaming algorithm with at least $0.8$-approximation
for each checkpoint. Unfortunately, \cite{DBLP:journals/mp/NemhauserWF78,DBLP:journals/jacm/Feige98} show that
there is no polynomial algorithm for \pName that can achieve an approximation ratio
of better than $1-\frac{1}{e} \approx 0.63$ unless P=NP.
Therefore, we devise a novel strategy to maintain
an adequate sequence of checkpoints so that
(1) the number of checkpoints is as few as possible for high efficiency;
(2) the utilities of the solutions still achieve a bounded approximation
ratio to the optimal one.

Towards both objectives, we propose the following strategy to maintain the checkpoints
in \histogram:
(1) create a checkpoint and a $\shortKs$ instance for each arrival element;
(2) delete a checkpoint and terminate its $\shortKs$ instance
once it can be approximated by any successive checkpoint.
Let $f[x_i,t]$ denote the utility of the solution
returned by $\mathcal{H}(x_{i})$ at time $t$.
Given three neighboring checkpoints $x_{i},x_{i+1},x_{i+2}$ ($i \in[1,s-2]$)
and a parameter $\beta > 0$,
if $f[x_{i+2},t] \geq (1-\beta) f[x_{i},t]$,
we consider the second checkpoint $x_{i+1}$ can be approximated
by the third one $x_{i+2}$.
In this case, $x_{i+1}$ will be deleted from \histogram.
We will formally analyze the soundness of such a strategy in Section~\ref{subsec:kwplus:analysis}.

\textbf{Buffer Maintenance and Post-Processing.}
To further improve the empirical performance of $\shortWinOpt$,
we maintain buffers along with the candidates in $\shortKs$ instances
and use these buffers for post-processing before returning the final solution.
The reasons why the buffers and post-processing are essential are as follows.
First, by using \histogram, the solutions of $\mathcal{H}(x_2)$ are always
used for $A_t$ when $t>W$. As it is common that $x_2 \gg t'$,
all elements between $v_{t'}$ and $v_{x_2-1}$
are missing from the solutions of $\mathcal{H}(x_2)$.
Second, the candidates with high thresholds in $\shortKs$ instances are hard to be filled,
even if more elements could still be added without violating the $d$-knapsack constraint.
Therefore, we maintain the buffers for post-processing
to improve the solution quality of $\shortWinOpt$.

We consider a buffer $B_{\phi}=\emptyset$ is initialized when each candidate $S_{\phi}$
in a $\shortKs$ instance $\mathcal{H}(x_i)$ ($i\in[1,s]$) is created.
When processing an element $v_t$, if adding $v_t$ to $S_{\phi}$ achieves
a marginal gain of slightly lower than the threshold, i.e.,
$\Delta_f(v_t|S_{\phi}) \geq \alpha \cdot \frac{\delta_t\cdot\phi}{1+d}$,
$v_t$ will be added to $B_{\phi}$.
Here, $\alpha\in(0,1)$ is used to control the lower bound for an element to be added to $B_{\phi}$.
Furthermore, we restrict the buffer size to $\eta$.
When the number of elements in $B_{\phi}$ exceeds $\eta$,
we first drop each element $v$ if $S_{\phi}\cup\{v\}\notin\xi$.
Then, we drop the elements with the least cost-effectivenesses w.r.t.~$S_{\phi}$
until $|B_{\phi}|=\eta$, where the cost-effectiveness of element $v$ is computed
by $\frac{\Delta_f(v|S_{\phi})}{\delta(v)}$, $\delta(v)=\max_{j\in[d]}c_j(v)$.
Before returning the solution at time $t$,
we perform the post-processing procedure using buffers
of $\mathcal{H}(x_1)$ and $\mathcal{H}(x_2)$
(if $t<W$, only the buffers of $\mathcal{H}(x_1)$ is used).
Specifically, for each candidate $S_{\phi}$, we run \algDen~\cite{DBLP:conf/acl/LinB11}
to add elements in buffers to $S_{\phi}$.
After post-processing each candidate,
we also return the candidate with the maximum utility as the final solution.

\begin{algorithm}
\begin{algorithmic}[1]
  \Require A stream $V=\langle v_1,v_2,\ldots \rangle$, the window size $W$, the buffer size $\eta$, the parameters $\alpha$ and $\beta$
  \Ensure The solution $S_t$ for \pName at time $t$
  \State Initialize $s \gets 0$, $X_0 \gets \emptyset$
  \For{$t \gets 1,2,\ldots$}
    \State $s \gets s+1$, $x_s \gets t$, and $X_t \gets X_{t-1} \circ \langle x_s \rangle$
    \label{line:kwplus:maintain:start}
    \State Initiate a $\shortKs$ instance $\mathcal{H}(x_s)$
    \State \textbf{for all} $S_{\phi}$ of $\mathcal{H}(x_s)$ \textbf{do} Initialize a buffer $B_{\phi} \gets \emptyset$
    \While{$t>W \wedge x_2<t'$}
    \label{line:kwplus:expire:start}
      \State $X_t \gets X_t \setminus \langle x_1 \rangle$, terminate $\mathcal{H}(x_1)$
      \State Shift the remaining checkpoints, $s \gets s-1$
    \EndWhile
    \label{line:kwplus:expire:end}
    \For{$i \gets 1,\ldots,s$}
    \label{line:kwplus:process:start}
      \State $\mathcal{H}(x_i)$ processes $v_t$ according to Algorithm~\ref{alg:ks}
      \State \textit{// buffer maintenance}
      \State \textbf{for all} $S_{\phi}$ of $\mathcal{H}(x_i)$ \textbf{do} \textsc{Buffer}($S_{\phi},B_{\phi},v_t$)
    \EndFor
    \label{line:kwplus:process:end}
    \While{$\exists i\in[1,s-2]$ : $f[x_{i+2},t] \geq (1-\beta) f[x_i,t]$}
    \label{line:kwplus:delete:start}
      \State $X_t \gets X_t \setminus \langle x_{i+1} \rangle$, terminate $\mathcal{H}(x_{i+1})$
      \State Shift the remaining checkpoints, $s\gets s-1$
    \EndWhile
    \label{line:kwplus:delete:end}
    \label{line:kwplus:maintain:end}
    \State \textit{// post-processing}
    \If{$x_1 \geq t'$}
    \label{line:kwplus:post:start}
      \State \textbf{for all} $S_{\phi}$ of $\mathcal{H}(x_1)$ \textbf{do} \textsc{CostEffectGreedy}($S_{\phi},B_{\phi}$)
      \State \Return $S_t \gets$ the result of $\mathcal{H}(x_1)$
    \Else
      \ForAll{$S_{\phi}$ of $\mathcal{H}(x_2)$}
        \State Add each element $v$ in $S'_{\phi}$ and $B'_{\phi}$ of $\mathcal{H}(x_1)$ to $B_{\phi}$ if $v$ does not expire and $S_{\phi} \cup \{v\} \in \xi$
        \State \textsc{CostEffectGreedy}($S_{\phi}, B_{\phi}$)
      \EndFor
      \State \Return $S_t \gets$ the result of $\mathcal{H}(x_2)$
    \EndIf
    \label{line:kwplus:post:end}
  \EndFor
  \Procedure{\textsc{Buffer}}{$S_{\phi},B_{\phi},v_t$}
  \label{line:kwplus:buffer:start}
    \If{$v_t \notin S_{\phi} \wedge \Delta_f(v_t|S_{\phi}) \geq \alpha \cdot \frac{\delta_t \cdot \phi}{1+d}$}
      \State $B_{\phi} \gets B_{\phi} \cup \{v_t\}$
    \EndIf
    \While{$|B_{\phi}|>\eta$}
      \State \textbf{for all} $v \in B_{\phi}$ \textbf{do} $B_{\phi} \gets B_{\phi} \setminus \{v\}$
             if $S_{\phi} \cup \{v\} \notin \xi$
      \State $v' \gets \argmin_{v \in B_{\phi}} \frac{\Delta_{f}(v|S_{\phi})}{\delta(v)}, \delta(v)=\max_{j\in[d]}c_j(v)$
      \State $B_{\phi} \gets B_{\phi} \setminus \{v'\}$
    \EndWhile
  \EndProcedure
  \label{line:kwplus:buffer:end}
  \Procedure{\textsc{CostEffectGreedy}}{$S_{\phi},B_{\phi}$}
  \label{line:kwplus:greedy:start}
    \While{$\exists v \in B_{\phi} : S_{\phi} \cup \{v\} \in \xi$}
      \State $v^{*} \gets \argmax_{v \in B_{\phi} \wedge S_{\phi} \cup \{v\} \in \xi} \frac{\Delta_f(v|S_{\phi})}{\delta(v)}$
      \State $S_{\phi} \gets S_{\phi} \cup \{v^{*}\}$, $B_{\phi} \gets B_{\phi} \setminus \{v^{*}\}$
    \EndWhile
  \EndProcedure
  \label{line:kwplus:greedy:end}
\end{algorithmic}
\caption{\algWindowOpt}\label{alg:kwplus}
\end{algorithm}

\textbf{Algorithmic Description.}
The pseudo-code of $\shortWinOpt$ is presented in Algorithm~\ref{alg:kwplus}.
The maintenance of \histogram is shown in Lines~\ref{line:kwplus:maintain:start}--\ref{line:kwplus:maintain:end}.
At time $t$, a new checkpoint $x_s=t$ and a $\shortKs$ instance $\mathcal{H}(x_s)$ are created for $v_t$.
Then, if there is more than one expired checkpoint in \histogram,
all except the last one will be deleted
(Lines~\ref{line:kwplus:expire:start}--\ref{line:kwplus:expire:end}).
This guarantees that there is only one expired checkpoint in $X_t$.
Subsequently, for each checkpoint $x_i$,
$\mathcal{H}(x_i)$ processes $v_t$ and updates the candidates independently
according to Lines~\ref{line:ks:stream:start}--\ref{line:ks:stream:end} of Algorithm~\ref{alg:ks}.
After updating the candidates of $\mathcal{H}(x_i)$ for $v_t$,
it performs the buffer maintenance procedure as follows
(Lines~\ref{line:kwplus:buffer:start}--\ref{line:kwplus:buffer:end}).
If $\Delta_f(v_t|S_{\phi}) \geq \alpha \cdot \frac{\delta_t\cdot\phi}{1+d}$,
$v_t$ is added to $B_{\phi}$.
When the number of elements in $B_{\phi}$ exceeds $\eta$,
it first drops any $v \in B_{\phi}$ if $S_{\phi}\cup\{v\}\notin\xi$
and then drops the element $v'$ with the least cost-effectiveness in $B_{\phi}$
until $|B_{\phi}|=\eta$.
Next, it maintains the checkpoints in \histogram.
The checkpoints that can be approximated by its successor
are identified and deleted from \histogram
(Lines~\ref{line:kwplus:delete:start}--\ref{line:kwplus:delete:end}).
After the \histogram maintenance, for any $x \in X_t$,
there is at most one checkpoint $x' \in X_t$ such that $x'>x$ and
$f[x',t] \geq (1-\beta) f[x,t]$.
Finally, the post-processing procedure is executed before returning
the solution $S_t$ for \pName w.r.t.~$A_t$.
When $t<W$, $\mathcal{H}(x_1)$ will provide $S_t$.
Each candidate $S_{\phi}$ in $\mathcal{H}(x_1)$ considers $B_{\phi}$ for post-processing.
Otherwise, $\mathcal{H}(x_2)$ will provide $S_t$.
We first add the non-expired elements in $S'_{\phi}$ and $B'_{\phi}$ of $\mathcal{H}(x_1)$ to $B_{\phi}$
for post-processing.
Starting from $S_{\phi}$, the post-processing procedure greedily adds the element $v^{*}$
with the maximum cost-effectiveness in $B_{\phi}$ to $S_{\phi}$
until none of the remaining elements in $B_{\phi}$ can be included without violating the $d$-knapsack constraint
(Lines~\ref{line:kwplus:greedy:start}--\ref{line:kwplus:greedy:end}).
After the post-processing, it also returns the candidate with the maximum utility
among the candidates in $\mathcal{H}(x_1)$ or $\mathcal{H}(x_2)$ as the final solution $S_t$.

\subsection{Theoretical Analysis}\label{subsec:kwplus:analysis}

Next, we analyze the approximation ratio and complexity of $\shortWinOpt$.
We first prove the properties of the checkpoints in \histogram (Lemma~\ref{lm:property}).
Based on the properties, we prove the approximation ratio of $\shortWinOpt$ (Theorem~\ref{thm:kwplusApprox}).
Finally, we analyze the number of checkpoints in \histogram,
calculate the cost of buffer maintenance and post-processing for $\shortWinOpt$,
and acquire the complexity of $\shortWinOpt$.

First of all, we prove the properties of the checkpoints in \histogram.
\begin{lemma}\label{lm:property}
  Given a parameter $\beta \in (0,1)$,
  each checkpoint $x_i \in X_t$ where $i\in[1,s]$ maintained by \histogram at time $t$
  satisfies one of the following properties:
  \begin{enumerate}
    \item if $f[x_{i+1},t] \geq (1-\beta) f[x_i,t]$, $f[x_{i+2},t] < (1-\beta) f[x_i,t]$ or $x_{i+1}=x_s$.
    \item if $x_{i+1} \neq x_i + 1$ and $f[x_{i+1},t] < (1-\beta) f[x_i,t]$, there exists some $t'<t$ such that $f[x_{i+1},t'] \geq (1-\beta) f[x_i,t']$.
    \item $x_{i+1} = x_i + 1$ and $f[x_{i+1},t] < (1-\beta) f[x_i,t]$.
  \end{enumerate}
\end{lemma}
\begin{proof}
  We prove the lemma by induction on $t$.
  As the base case, we first check the condition when $t=2$ and $X_2=\langle x_1=1 ,x_2=2 \rangle$.
  Then, Property (1) holds if $f[x_2,2] \geq (1-\beta) f[x_1,2]$;
  otherwise, Property (3) holds.

  Next, we assume Lemma~\ref{lm:property} holds at time $t$ and show that it still holds
  after performing Lines~\ref{line:kwplus:maintain:start}--\ref{line:kwplus:maintain:end}
  of Algorithm~\ref{alg:kwplus} at time $t+1$.
  Let $x_i$ be a checkpoint that is created before $t+1$ and not deleted
  during the maintenance at time $t+1$ and
  $x_{i+1}$ be the checkpoint next to $x_i$ at time $t$.
  We discuss all possible cases during the maintenance at time $t+1$.

  \textbf{Case 1}.
  $x_{i+1} \neq x_i + 1$ and $x_{i+1}$ is deleted from \histogram at time $t+1$.
  In this case, we have $f[x_{i+2},t+1] \geq (1-\beta) f[x_i,t+1]$
  (Line~\ref{line:kwplus:delete:start} of Algorithm~\ref{alg:kwplus}).
  As $x_{i+2}$ becomes the successor of $x_i$ at time $t+1$, Property (1) holds.

  \textbf{Case 2}.
  $x_{i+1} \neq x_i + 1$ and $x_{i+1}$ is not deleted from \histogram at time $t+1$.
  In this case, we consider $x_{i+1}$ becomes the successor of $x_i$ at some time $t' \leq t$.
  Then, it must hold that $f[x_{i+1},t'] \geq (1-\beta) f[x_i,t']$.
  Since $x_{i+1}$ is not deleted at time $t+1$, either Property (1)
  (if $f[x_{i+1},t+1] \geq (1-\beta) f[x_i,t+1]$)
  or Property (2) (if $f[x_{i+1},t+1] < (1-\beta) f[x_i,t+1]$) holds.

  \textbf{Case 3}.
  $x_{i+1} = x_i + 1$. No matter whether $x_{i+1}$ is deleted
  at time $t+1$, Property (1) holds if $f[x_{i+1},t+1] \geq (1-\beta) f[x_i,t+1]$;
  otherwise, Property (3) holds.

  We show that the properties of \histogram still hold at time $t+1$
  in all possible cases and conclude the proof.
\end{proof}

Given the properties of \histogram,
we can analyze the approximation ratio
of $S_t$ returned by Algorithm~\ref{alg:kwplus} for \pName w.r.t.~$A_t$.

\begin{theorem}\label{thm:kwplusApprox}
  The solution $S_t$ returned by Algorithm~\ref{alg:kwplus} satisfies that
  $f(S_t) \geq \frac{1-\varepsilon'}{2(1+d)}\cdot\optimal_t$ at any time $t$
  where $\varepsilon'=\varepsilon+\beta$.
\end{theorem}
\begin{proof}
  We consider the first two checkpoints $x_1$ and $x_2$ of \histogram at time $t$
  and assume that post-processing does not change the solution $S_t$.
  If $t \leq W$, $x_1 = 1$ does not expire
  and $\mathcal{H}(x_1)$ are maintained over $A_t= \langle v_1,\ldots,v_t \rangle$.
  Thus, $f(S_t) = f[x_1,t] \geq \frac{1-\varepsilon}{1+d}\optimal_t$ for $t \leq W$
  by Theorem~\ref{thm:ksApprox}.
  Next, we consider $t>W$ and $x_2=x_1+1$.
  In this case, $x_1$ expires and $x_2$ corresponds to the starting point of $A_t$.
  Similarly, $f(S_t) = f[x_2,t] \geq \frac{1-\varepsilon}{1+d} \optimal_t$.

  Subsequently, we consider other cases for $t>W$.
  We use $\optimal^{x}_{y}$ to denote the optimal utility of \pName w.r.t. the elements
  $\langle v_{x},\ldots,v_{y} \rangle$.

  \textbf{Case 1}.
  If $f[x_2,t] \geq (1-\beta) f[x_1,t]$, $f(S_t) = f[x_2,t] \geq (1-\beta) f[x_1,t]$.
  By Theorem~\ref{thm:ksApprox}, $f[x_1,t] \geq \frac{1-\varepsilon}{1+d} \optimal^{x_1}_{t}$.
  As $x_1<t'$, we have $A_t \subset \langle v_{x_1},\ldots,v_t \rangle$
  and $\optimal_t \leq \optimal^{x_1}_{t}$.
  Finally, we have $f(S_t) \geq \frac{(1-\beta)(1-\varepsilon)}{1+d} \optimal_t$.

  \textbf{Case 2}.
  If $f[x_2,t] < (1-\beta) f[x_1,t]$,
  we have $f[x_2,t'] \geq (1-\beta) f[x_1,t']$ for some $t'<t$.
  Let $S^*_{x_1,t}$ denote the optimal solution for $\langle v_{x_1},\ldots,v_t \rangle$.
  We can split $S^*_{x_1,t}$ into two subsets $S_1$ and $S_2$,
  where $S_1=\{v_i|v_i \in S^*_{x_1,t} \wedge i \in [x_1,t']\}$
  and $S_2=\{v_i|v_i \in S^*_{x_1,t} \wedge i \in [x_2,t]\}$.
  Let $\optimal_1=f(S_1)$ and $\optimal_2=f(S_2)$.
  For $S^*_{x_1,t}=S_1 \cup S_2$ and the submodularity of $f(\cdot)$,
  $\optimal^{x_1}_{t} \leq \optimal_1 + \optimal_2$.
  Then, as $S_1\in\xi$ and $S_2\in\xi$,
  it holds that $\optimal_1 \leq \optimal^{x_1}_{t'}$
  and $\optimal_2 \leq \optimal^{x_2}_{t}$.
  In addition, for any $t_1<t_2$,
  the solution returned by \algStream satisfies that $f[x,t_1] \leq f[x,t_2]$.
  As $t>t'$, we have:
  \begin{displaymath}
    f[x_2,t] \geq \frac{(1-\beta)(1-\varepsilon)}{1+d}\cdot\optimal^{x_1}_{t'} \geq \frac{(1-\beta)(1-\varepsilon)}{1+d}\cdot\optimal_1
  \end{displaymath}
  We also have:
  \begin{displaymath}
    f[x_2,t] \geq \frac{1-\varepsilon}{1+d}\cdot\optimal^{x_2}_{t} \geq \frac{1-\varepsilon}{1+d}\cdot\optimal_2
  \end{displaymath}
  Adding the above two inequalities, we prove:
  \begin{equation}\label{eq:kwplus:ratio}
    f(S_t) = f[x_2,t] \geq \frac{(1-\beta)(1-\varepsilon)}{2(1+d)}\cdot\optimal_t
  \end{equation}

  Finally, because the post-processing procedure must not decrease the utility of any candidate,
  Equation~\ref{eq:kwplus:ratio} still holds after post-processing.
  Thus, we conclude the proof by replacing $\lambda$ and $\varepsilon$ with $\varepsilon'$
  in Equation~\ref{eq:kwplus:ratio}.
\end{proof}

\textbf{The Complexity of $\shortWinOpt$.}
According to Lemma~\ref{lm:property}, either $f[x_{i+1},t]$ or $f[x_{i+2},t]$
is less that $(1-\beta)f[x_i,t]$ at any time $t$.
Given $\theta=\frac{f[x_1,t]}{f[x_s,t]}$, the number of checkpoints in \histogram is
at most $\lceil\frac{2\log\theta}{\log(1-\beta)^{-1}}\rceil$.
Therefore, the number of checkpoints is $\mathcal{O}(\frac{\log\theta}{\beta})$.
$\shortWinOpt$ performs $\mathcal{O}(\frac{\log\theta\cdot\log(d\cdot\gamma^{-1})}{\varepsilon'^{2}})$
function calls to update the candidates in the checkpoints for one element
and stores at most $\mathcal{O}(\frac{\log\theta\cdot\log(d\cdot\gamma^{-1})}{\gamma\cdot\varepsilon'^{2}})$
elements within the candidates.
In practice, the buffer of each candidate is implemented by a min-heap
and the buffer size $\eta=\mathcal{O}(\gamma^{-1})$.
The complexity of adding an element to the buffer is $\mathcal{O}(\log\gamma^{-1})$
and dropping elements from the buffer is $\mathcal{O}(\gamma^{-1})$.
Thus, the amortized computational cost for buffer maintenance is
$\mathcal{O}(\frac{\log\theta\cdot\log(d\cdot\gamma^{-1})}{\varepsilon'^{2}})$
and the total number of elements in buffers is
$\mathcal{O}(\frac{\log\theta\cdot\log(d\cdot\gamma^{-1})}{\gamma\cdot\varepsilon'^{2}})$.
The post-processing for one candidate handles $\mathcal{O}(\gamma^{-1})$ elements
and runs at most $\gamma^{-1}$ iterations.
Therefore, the post-processing requires
$\mathcal{O}(\frac{\log(d\cdot\gamma^{-1})}{\gamma^{2}\cdot\varepsilon'})$ function calls.
Generally, $\shortWinOpt$ runs in
$\mathcal{O}\big(\frac{\log(d\cdot\gamma^{-1})}{\varepsilon'}\cdot(\gamma^{-2}+\frac{\log\theta}{\varepsilon'})\big)$
time to process one element
and stores $\mathcal{O}(\frac{\log\theta\cdot\log(d\cdot\gamma^{-1})}{\gamma\cdot\varepsilon'^{2}})$ elements in total.

\subsection{Discussion}
\label{subsec:discussion}

In practice, it is no need to update the solution for every arrival element.
The update is often performed in a batch manner.
Specifically, we consider the sliding window receives $T$ new elements
while the earliest $T$ elements become expired at time $t$.
Both $\shortWin$ and $\shortWinOpt$ can handle the scenario with trivial adaptations.
For $\shortWin$, it also stores the active elements in $A_t$ and creates a checkpoint
for every $L$ elements. The only difference is that the interval $L$ becomes
$\sqrt{W \cdot T}$ while the number of checkpoints $s$ decreases to $\sqrt{\frac{W}{T}}$.
For $\shortWinOpt$, it creates one checkpoint at each time $t$
and updates existing checkpoints by processing a batch of elements from $v_{t-T+1}$ to $v_{t}$ collectively.
In this way, the total number of checkpoints created is $\lceil\frac{W}{T}\rceil$.
The number of checkpoints in \histogram is determined by the utilities and thus is not affected.
In addition, any other theoretical results,
the buffer maintenance, and the post-processing procedure
are also not affected by these adaptations.

\section{Experiments}\label{sec:experiment}

In this section, we report our experimental results for two \rss applications
(as presented in Section~\ref{sec:application}) in real-world datasets.
First, we introduce the experimental setup in Section~\ref{subsec:settings}.
Then, we evaluate the effectiveness and efficiency of our proposed frameworks
compared with several baselines in Section~\ref{subsec:results}.

\subsection{Experimental Setup}\label{subsec:settings}

\textbf{Datasets.}
Two real-world datasets are used in our experiments.
First, we use the \emph{Twitter} dataset for \emph{social stream summarization} (see Section~\ref{subsec:social:stream}).
It is collected via the streaming API\footnote{\url{http://twitter4j.org/en/index.html}}
and contains $18,770,231$ tweets and $8,071,484$ words.
The average number of words in each tweet is $\overline{l}=4.8$
and the average number of followers of each user is $\overline{fl}=521.4$.
In the experiments, we feed each tweet to the compared approaches one by one
in ascending order of timestamp.
Second, we use the \emph{Yahoo! Webscope} dataset\footnote{\url{http://webscope.sandbox.yahoo.com}}
for \emph{active set selection} (see Section~\ref{subsec:active:set}).
It consists of $45,811,883$ user visits from the Featured Tab of the Today
module on the Yahoo! front page.
Each user visit is a $5$-dimensional feature vector.
We set $h=0.75$ and $\sigma=1$ in Equation~\ref{eq:IVM} following~\cite{DBLP:conf/kdd/BadanidiyuruMKK14}.
The costs are generated from a uniform distribution $\mathcal{U}(0.02,0.08)$.
In the experiments, we feed all user visits to the compared approaches one by one in the same order.

\textbf{Additional constraints.}
To evaluate the compared approaches with varying the dimension of knapsacks, i.e., $d$,
we generate additional constraints by assigning random costs to each element in both datasets.
Specifically, we generate a $5$-dimensional cost vector $\mathbf{c}(v)=\{c_1(v),\ldots,c_5(v)\}$
for each element $v$.
And each cost is generated independently from a uniform distribution $\mathcal{U}(0.02,0.08)$.
We set $d$ to range from $1$ to $5$ in the experiments and use
the first $d$ dimensions of $\mathbf{c}(v)$ for the $d$-knapsack constraint.

\begin{table}[t]
  \centering
  \caption{The parameters tested in the experiments}
  \label{tbl:parameters}
  \begin{tabular}{|c|l|}
    \hline
    \textbf{Parameter} & \textbf{Values} \\
    \hline
    $d$ & 1, 2, 3, 4, 5 \\
    \hline
    $c$ & 0.02, \textbf{0.04}, 0.06, 0.08, 0.1 \\
    \hline
    $W$ & 100k, \textbf{200k}, 300k, 400k, 500k \\
    \hline
    $\lambda$ & 0.05, \textbf{0.1}, 0.15, 0.2, 0.25 \\
    \hline
    $\beta$ & 0.05, \textbf{0.1}, 0.15, 0.2, 0.25 \\
    \hline
  \end{tabular}
\end{table}

\textbf{Compared Approaches.}
The approaches compared in our experiments are listed as follows.
\begin{itemize}
  \item \algDen ($\shortDen$).
  We implement the \algDen algorithm~\cite{DBLP:conf/kdd/LeskovecKGFVG07}
  as the batch baseline. Since $\shortDen$ is designed for submodular maximization
  with a $1$-knapsack constraint, we slightly adapt it
  for \pName: the cost-effectiveness of $v$ w.r.t.~$S$ is computed by
  $\frac{\Delta_{f}(v|S)}{\delta(v)}$ where $\delta(v)=\max_{j\in[d]}c(v)$.
  To work in the sliding window model,
  it stores the active elements in $A_t$ and recomputes
  the solution from scratch for each window slide.
  \item \textsc{Streaming} ($\shortStr$).
  We implement the state-of-the-art append-only streaming algorithm~\cite{DBLP:conf/globalsip/YuXC16}
  for \pName
  as a baseline. To work in the sliding window model,
  it also stores the active elements in $A_t$
  and recomputes the solution from scratch for each window slide.
  \item \textsc{Window} ($\shortCardWin$).
  We implement the state-of-the-art algorithm for submodular maximization
  over sliding windows~\cite{DBLP:conf/www/EpastoLVZ17} as a baseline.
  Since it only works with one cardinality constraint,
  we cast the $d$-knapsack constraint to the cardinality constraint
  by setting the budget $k=\frac{1}{\gamma}$
  where $\gamma$ is the average cost of elements.
  When maintaining the solutions over sliding windows,
  it only considers the marginal gains
  of elements and treats the cost of any element as $1$.
  \item \algWindow ($\shortWin$).
  We implement the \algWindow framework in Section~\ref{sec:framework}.
  \item \algWindowOpt ($\shortWinOpt$).
  We implement the \algWindowOpt framework in Section~\ref{sec:kwplus}.
  We set $\alpha=0.5$ and $\eta=20$ for buffer maintenance.
\end{itemize}

\textbf{Parameters.}
The parameters tested in our experiments are listed in Table~\ref{tbl:parameters}
with default values in bold.
$d$ is the dimension of the knapsack constraint.
We use $d=3$ for social stream summarization
and $d=1$ for active set selection by default as introduced in Section~\ref{sec:application};
$c$ is the average cost of each element.
For the \emph{Twitter} dataset, we set $k=\frac{1}{c}$ to assign the costs $c_1(v), c_2(v), c_3(v)$
accordingly as introduced in Section~\ref{subsec:social:stream}.
For the \emph{Yahoo! Webscope} dataset, the average of generated costs is $c=0.05$.
We scale the costs linearly in the experiments for varying $c$.
$W$ is the size of the sliding window.
We set the number of elements for each window slide to $T = 0.01\% \cdot W$.
The interval for neighboring checkpoints in $\shortWin$
is $L = \sqrt{W \cdot T} = 1\% \cdot W$ (Section~\ref{subsec:discussion}).
$\lambda$ is the parameter used in $\shortWin$, $\shortWinOpt$, $\shortStr$, and $\shortCardWin$
for the balance between the number of candidates maintained for processing append-only streams and solution quality.
$\beta$ is the parameter for $\shortWinOpt$ to balance
between the number of checkpoints and solution quality.

\textbf{Metrics.}
We consider the following metrics to evaluate the compared approaches.
\begin{itemize}
  \item \textbf{CPU time} is the average CPU time of an approach to process one window slide.
  It is used to measure the efficiency of compared approaches.
  \item \textbf{Utility} is the average utility value of
  the solution returned by an approach for each window.
  It evaluates the solution quality of compared approaches.
  \item \textbf{\#checkpoints} and \textbf{\#elements} are the average numbers of
  checkpoints and elements maintained by $\shortWinOpt$,
  which are used to measure its space usage.
\end{itemize}

\textbf{Experimental Environment.}
All the above approaches are implemented in Java 8
and the experiments are conducted on a server running Ubuntu 16.04
with a 1.9GHz Intel Xeon E7-4820 processor and 128 GB memory.

\subsection{Experimental Results}\label{subsec:results}

\begin{figure}
  \centering
  \subfloat[Twitter]{
    \includegraphics[width=0.3\textwidth]{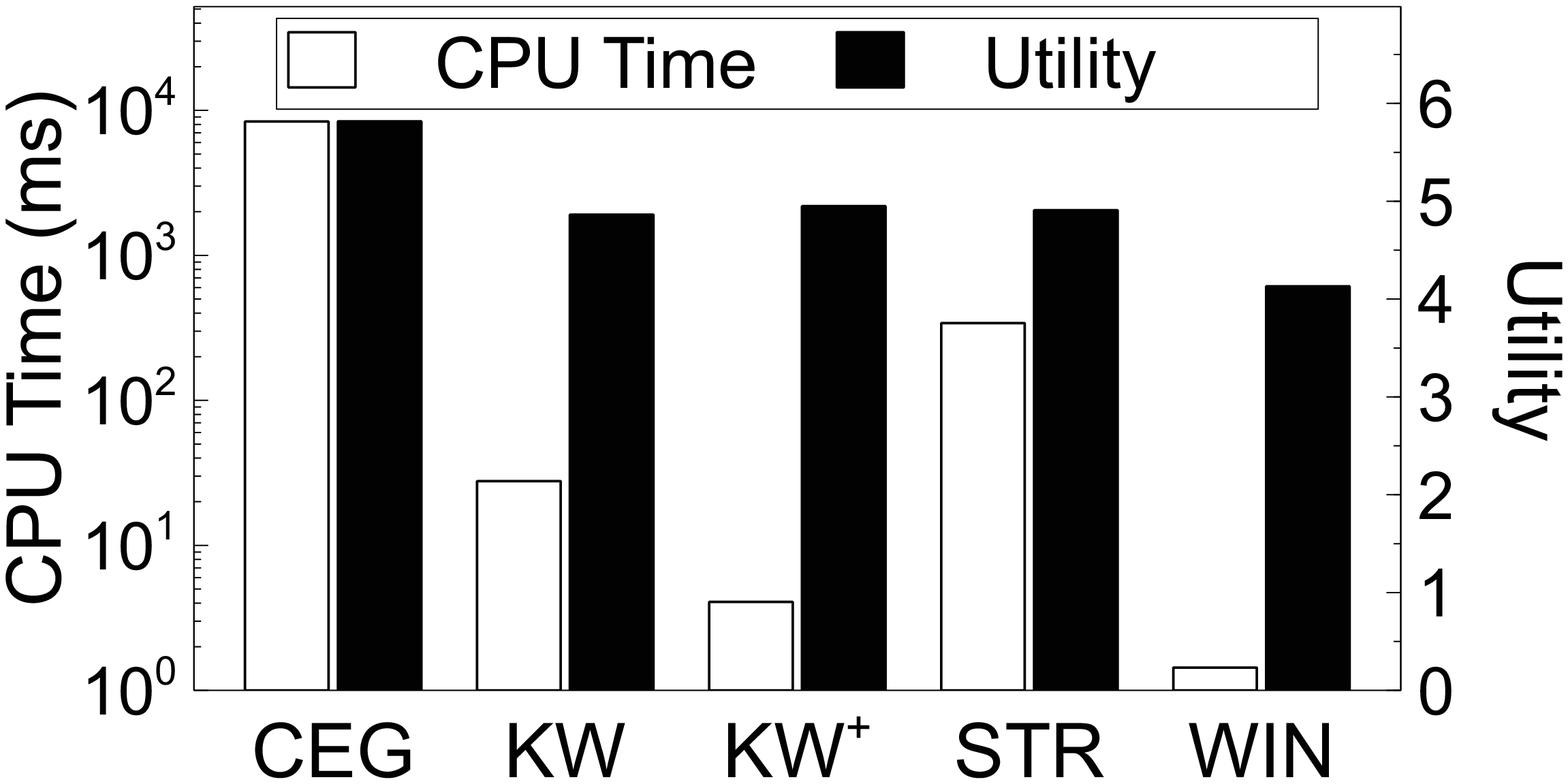}
  }
  \subfloat[Yahoo! Webscope]{
    \includegraphics[width=0.3\textwidth]{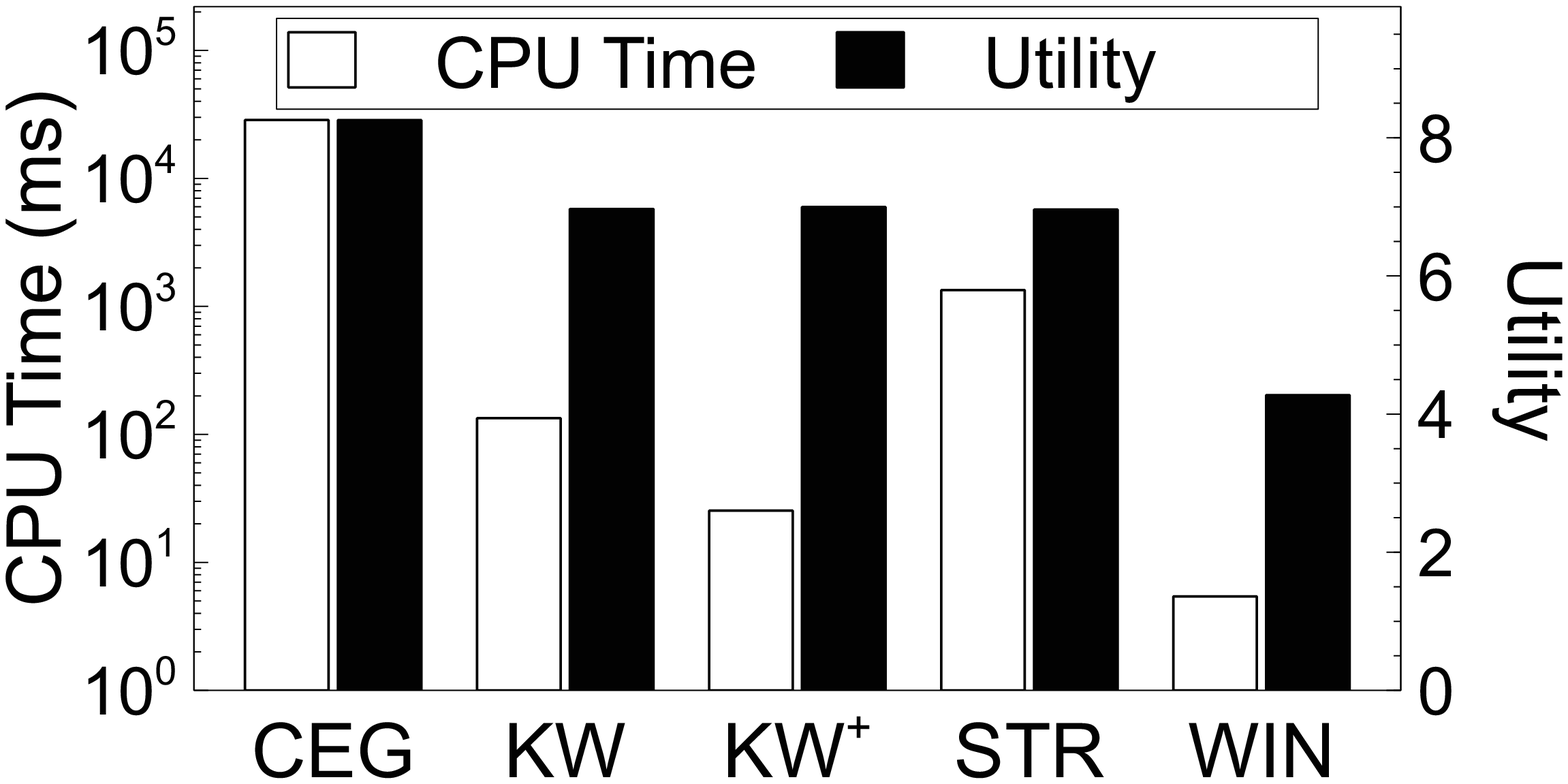}
  }
  \caption{The overall experimental results.}
  \label{fig:time-utility-default}
\end{figure}

\textbf{Overall Results.}
In Figure~\ref{fig:time-utility-default}, we present the \emph{CPU time}
and \emph{utilities} of compared approaches in the default setting.
Although $\shortDen$ achieves the best utilities,
it takes around 10s to process each window slide,
which is far lower than the rates of real-world data streams.
$\shortWin$ and $\shortWinOpt$ run over two and three orders of magnitude
faster than $\shortDen$ respectively and can process each window slide within 100ms.
Meanwhile, the utilities of the solutions provided by $\shortWin$ and $\shortWinOpt$
are about $85\%$ of those of $\shortDen$.
Furthermore, $\shortWinOpt$ significantly improves the efficiency upon $\shortWin$,
achieving speedups of at least $6$x in both datasets.
Compared with $\shortStr$, $\shortWin$ and $\shortWinOpt$ run dozens of times faster
while providing solutions with similar utilities.
Finally, we observe $\shortCardWin$ runs faster than other approaches
but shows obviously inferior solution quality.
This is because $\shortCardWin$ treats the costs of any element equally
and only considers marginal utility gains when adding an element.
As a result, the solutions of $\shortCardWin$ contain fewer elements than other approaches,
which leads to both higher efficiency and worse solution quality.

\begin{figure}[t]
    \centering
    \includegraphics[width=\textwidth]{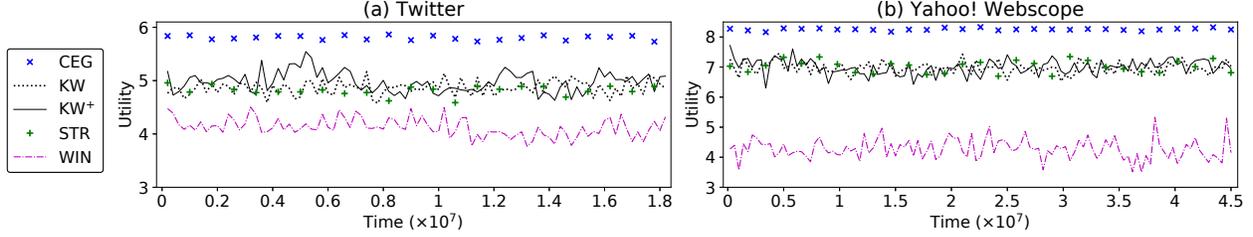}
    \vspace{-2em}
    \caption{The utilities of compared approaches over time.
    Note that we retrieve the solutions of $\shortDen$ and $\shortStr$ only at sampled timestamps.
    The solutions of $\shortWin$, $\shortWinOpt$, and $\shortCardWin$ are returned for every window slide.}
    \label{fig:window-evolve}
\end{figure}

In Figure~\ref{fig:window-evolve}, we present the utilities of compared approaches
from time $t=W$ to the end of the stream $t=n$.
The solutions returned by $\shortDen$ achieve the highest utilities all the time.
the solution utilities of $\shortWin$, $\shortWinOpt$, and $\shortStr$ fluctuate
over time and are generally close to each other.
But remember that $\shortWinOpt$ takes much less CPU time than $\shortWin$
while $\shortWin$ runs significantly faster than $\shortStr$
(as illustrated in Figure~\ref{fig:time-utility-default}).
Also as expected, the solution quality of $\shortCardWin$
cannot match any other approaches.

To sum up,
$\shortWinOpt$ achieves the best balance between efficiency and solution quality:
compared with $\shortDen$, it runs more than three orders of magnitude faster while providing
solutions with 85\% average utility;
it has much higher efficiency than $\shortWin$ and $\shortStr$ but achieves equivalent solution quality;
it significantly improves the solution quality upon $\shortCardWin$ at a little expense of efficiency.

\begin{figure}[t]
    \centering
    \includegraphics[width=\textwidth]{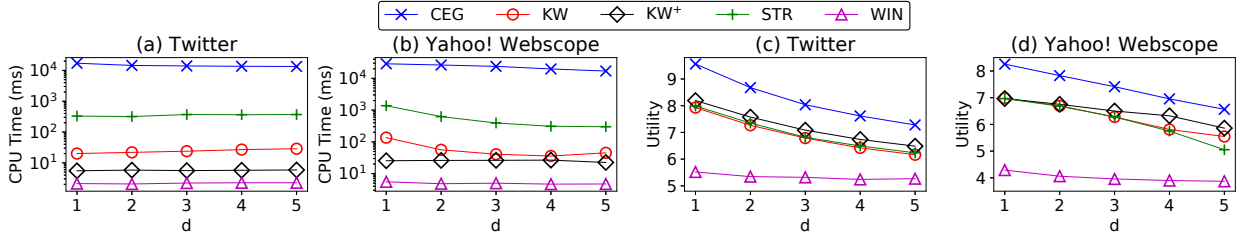}
    \vspace{-2em}
    \caption{The CPU time and utilities of compared approaches with varying the dimension $d$ of the knapsack constraint.}
    \label{fig:varying-d}
\end{figure}

\textbf{Effect of $d$.}
The CPU time and utilities of compared approaches with varying $d$ are
shown in Figure~\ref{fig:varying-d}. The CPU time of $\shortDen$ decreases
when $d$ increases. This is because the average solution size becomes
smaller when there are more constraints. The CPU time of $\shortWin$
shows different trends in both datasets: it decreases in the
\emph{Yahoo! Webscope} dataset but keeps steady in the \emph{Twitter}
dataset when $d$ becomes larger. There are two observations behind
such trends: First, since $\shortKs$ maintains the candidates for
estimations from $m$ to $M(1+d)$ (see Algorithm~\ref{alg:ks}),
a $\shortKs$ instance maintains more candidates with increasing $d$.
Second, the average solution size decreases with $d$.
In the \emph{Twitter} dataset, the extra costs for maintaining more candidates
cancel out the benefits of smaller solutions and thus the overall CPU time keeps steady.
However, in the \emph{Yahoo! Webscope} dataset,
the time complexity of evaluating IVM in Equation~\ref{eq:IVM}
for a set $S$ is $\mathcal{O}(|S|^3)$.
As a result, the CPU time for each IVM evaluation is very sensitive to $|S|$.
Although more candidates are maintained, the overall CPU time of $\shortWin$
still becomes much lower. The CPU time of $\shortWinOpt$ shows a similar
trend to $\shortWin$ in the \emph{Twitter} dataset.
But it keeps steady with increasing $d$ in the \emph{Yahoo! Webscope} dataset.
The reason behind such an observation is, although the CPU time to update the checkpoints
decreases, the post-processing takes longer time when $d$ increases.
The utilities of all compared approaches decrease when $d$ increases
because of smaller solution sizes.
Compared with $\shortWin$ and $\shortStr$, $\shortWinOpt$ shows
slightly better solution quality for a larger $d$
due to the benefits of post-processing.
In addition, the ratios between the utilities of the solutions of
$\shortStr$, $\shortWin$, and $\shortWinOpt$ and those of $\shortDen$
are $84\%$--$90\%$ and remain stable for different $d$.

\begin{figure}[t]
    \centering
    \includegraphics[width=\textwidth]{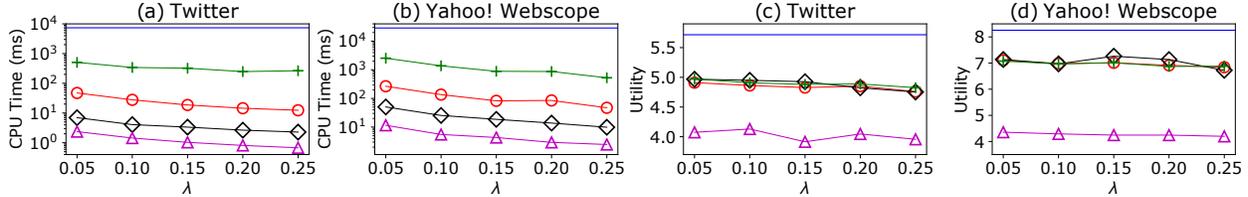}
    \vspace{-2em}
    \caption{The CPU time and utilities of compared approaches with varying the parameter $\lambda$. Note that $\shortDen$ is not affected by $\lambda$. We use horizontal blue lines to represent the CPU time and utilities of $\shortDen$ for ease of comparison.}
    \label{fig:varying-lambda}
\end{figure}

\begin{figure}[t]
    \centering
    \includegraphics[width=\textwidth]{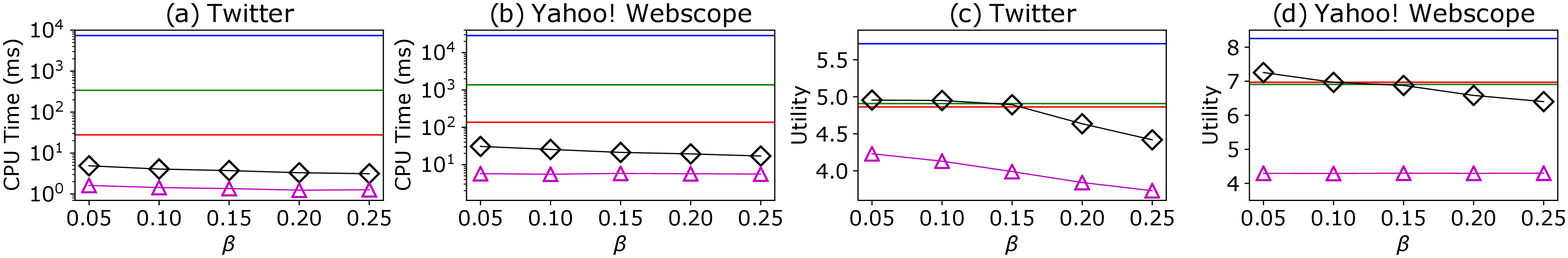}
    \vspace{-2em}
    \caption{The CPU time and utilities of compared approaches with varying the parameter $\beta$. Note that $\shortDen$, $\shortWin$, and $\shortStr$ are not affected by $\beta$. For ease of comparison, we use horizontal blue, red, and green lines to represent the CPU time and utilities of $\shortDen$, $\shortWin$, and $\shortStr$ respectively.}
    \label{fig:varying-beta}
\end{figure}

\textbf{Robustness against $\lambda$ and $\beta$.}
The experimental results of compared approaches with varying parameters $\lambda$
are shown in Figure~\ref{fig:varying-lambda}.
For all compared approaches except $\shortDen$, the CPU time obviously drops
with increasing $\lambda$. This is because the number of candidates is inversely correlated
to $\lambda$. However, we observe that their utilities are rather robust against $\lambda$
and only slightly decrease for a larger $\lambda$. The utility of $\shortWinOpt$
in the \emph{Yahoo! Webscope} dataset even increases when $\lambda=0.15$ thanks to the post-processing.

The experimental results of compared approaches with varying parameters $\beta$
are shown in Figure~\ref{fig:varying-beta}.
Because the number of checkpoints and $\shortKs$ instances in \histogram
is inversely correlated to $\beta$,
the CPU time of $\shortWinOpt$ decreases when $\beta$ increases.
However, the robustness of $\shortWinOpt$ against $\beta$ is worse than
its robustness against $\lambda$.
The utilities show drastic drops when $\beta=0.2$ or $0.25$.
As the intervals between the first two checkpoints increase with $\beta$,
the errors of using the results from the second checkpoint
as the solutions inevitably increase.
Considering the results, we advise using a small $\beta$
so that $\shortWinOpt$ can achieve good solution quality.

\begin{figure}[t]
  \centering
  \includegraphics[width=\textwidth]{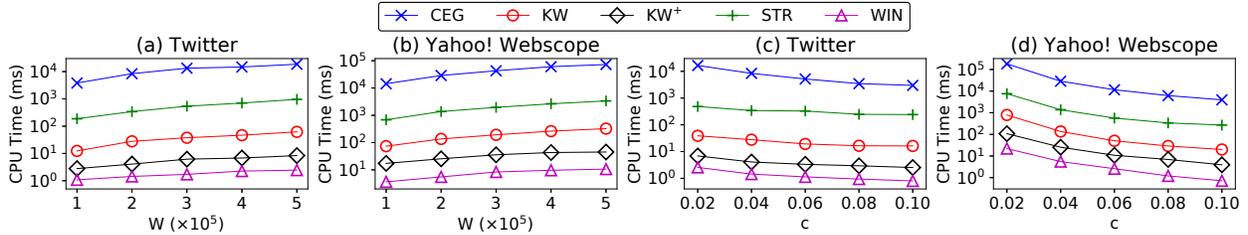}
  \vspace{-2em}
  \caption{The CPU time of compared approaches with varying the window size $W$ and the average cost $c$.}
  \label{fig:varying-w-c}
\end{figure}

\begin{table}[t]
  \centering
  \footnotesize
  \setlength{\tabcolsep}{2.5pt}
  \caption{The number of checkpoints and elements (including candidates and buffers) maintained by $\shortWinOpt$.}
  \label{tbl:checkpoints}
  \begin{tabular}{|l|l||c|c|c|c|c|c|c|c|c|c|}
    \hline
    \multirow{2}{*}{\textbf{Dataset}} & \textbf{Parameter} & \multicolumn{5}{c|}{$W$} & \multicolumn{5}{c|}{$c$} \\ \cline{2-12}
    & \textbf{Value} & 100k & 200k & 300k & 400k & 500k & 0.02 & 0.04 & 0.06 & 0.08 & 0.1 \\ \hline\hline
    \multirow{2}{*}{Twitter} & \#checkpoints & 4.89 & 4.49 & 4.32 & 4.34 & 4.27
    & 4.44 & 4.49 & 4.15 & 4.42 & 4.68 \\ \cline{2-12}
    & \#elements & 3949.9 & 3770.4 & 3674.2 & 3725.3 & 3676.1
    & 5489.4 & 3770.4 & 2843.4 & 2618.1 & 2509.5 \\ \hline\hline
    \multirow{2}{*}{Yahoo! Webscope} & \#checkpoints & 4.44 & 3.58 & 3.62 & 3.5 & 2.8
    & 4.4 & 3.58 & 3.18 & 3.44 & 2.7 \\ \cline{2-12}
    & \#elements & 3258.6 & 2617.54 & 2735.38 & 2647.16 & 2019.9
    & 6036.86 & 2617.54 & 1680.88 & 1454.92 & 908.16 \\ \hline
  \end{tabular}
\end{table}

\textbf{Scalability.}
In Figure~\ref{fig:varying-w-c}, we present the CPU time of compared approaches
with varying $W$ and $c$.
The CPU time to process each window slide increases with $W$.
This is because the number of elements processed for each window slide
is set to $0.01\% \cdot W$ which increases linearly with $W$.
For all compared approaches, it takes a longer CPU time when $c$ decreases
because the solution size is inversely proportional to $c$.
In the \emph{Yahoo! Webscope} dataset, the CPU time increases
drastically when $c$ decreases because the time complexity of the IVM function evaluation is $\mathcal{O}(|S|^3)$.
Thus, all compared approaches spend much more CPU time for each evaluation of $f(S)$
when the solution size grows.
Nevertheless, the CPU time of $\shortWinOpt$ and that of $\shortWin$ are within $100$ms and $1$s
respectively in all parameter settings.

We list the number of checkpoints and the number of elements
in both candidates and buffers maintained by $\shortWinOpt$
with varying $W$ and $c$ in Table~\ref{tbl:checkpoints}.
First, because the number of checkpoints and the number of elements
are independent of $W$ and bounded by the ratio of the utilities of
the solutions provided by the first and last checkpoints,
both metrics hardly increases with $W$.
In addition, the number of elements in $\shortWinOpt$ increases when $c$ decreases
because each candidate maintains more elements.
Generally, $\shortWinOpt$ only stores several thousand elements
when $W$ ranges from $100$k to $500$k. Taking $W=500$k as an example,
$\shortWinOpt$ merely stores $0.7\%$ of the active elements.
Therefore, the space usage of $\shortWinOpt$ is much smaller than
$\shortWin$, $\shortDen$, and $\shortStr$,
which need to store the entire active window.
Furthermore, the number of elements maintained by $\shortWinOpt$
does not increase with the window size $W$
because the space complexity of $\shortWinOpt$ is independent of $W$.
Hence, $\shortWinOpt$ is scalable for large window sizes.

\section{Related Work}\label{sec:related}

\begin{table}
  \centering
  \footnotesize
  \setlength{\tabcolsep}{2pt}
  \caption{A theoretical comparison of existing submodular maximization algorithms.
  The algorithms proposed in this work are highlighted by $^*$.}
  \label{tbl:comparison}
  \begin{tabular}{|l|c|c|c|c|}
    \hline
    \textbf{Algorithm} & \textbf{Data model} & \textbf{Constraint} & \textbf{Approximation} & \textbf{Time complexity} \\\hline
    Sviridenko~\cite{DBLP:journals/orl/Sviridenko04} & batch & $1$-knapsack & $1-\frac{1}{e}$ & $\mathcal{O}(W^5)$ \\\hline
    Kulik et al.~\cite{DBLP:conf/soda/KulikST09} & batch & $d$-knapsack & $1-\frac{1}{e}-\varepsilon$ & $\mathcal{O}(W^{d\cdot\varepsilon^{-4}})$ \\\hline
    Badanidiyuru et al.~\cite{DBLP:conf/soda/BadanidiyuruV14} & batch & $1$-knapsack & $1-\frac{1}{e}-\varepsilon$
    & $\mathcal{O}(W^{2}\cdot(\varepsilon^{-1} \cdot \log W)^{\varepsilon^{-8}})$ \\\hline
    Leskovec et al.~\cite{DBLP:conf/kdd/LeskovecKGFVG07} \& Lin et al.~\cite{DBLP:conf/acl/LinB11} & batch & $1$-knapsack & $\frac{1}{2}(1-\frac{1}{e})$ & $\mathcal{O}(\gamma^{-1} \cdot W)$ \\\hline
    Badanidiyuru et al.~\cite{DBLP:conf/kdd/BadanidiyuruMKK14} \& Kumar et al.~\cite{DBLP:journals/topc/KumarMVV15} & append-only stream & cardinality & $\frac{1}{2}-\varepsilon$ & $\mathcal{O}(\frac{\log k}{\varepsilon})$ \\\hline
    Huang et al.~\cite{DBLP:conf/approx/HuangKY17} & append-only stream & $1$-knapsack & $\frac{4}{11}-\varepsilon$ & $\mathcal{O}((\frac{\log \gamma^{-1}}{\varepsilon})^{4})$ \\\hline
    Yu et al.~\cite{DBLP:conf/globalsip/YuXC16} & append-only stream & $d$-knapsack & $\frac{1}{1+2d}-\varepsilon$ & $\mathcal{O}(\frac{\log \gamma^{-1}}{\varepsilon})$ \\\hline
    Epasto et al.~\cite{DBLP:conf/www/EpastoLVZ17} & sliding window & cardinality & $\frac{1}{3}-\varepsilon$ & $\mathcal{O}(\frac{\log^{2} (k\cdot\theta)}{\varepsilon^2})$ \\\hline
    \hline
    \textbf{\algStream ($\shortKs$)$^{*}$} & append-only stream & $d$-knapsack & $\frac{1-\varepsilon}{1+d}$ & $\mathcal{O}(\frac{\log(d \cdot \gamma^{-1})}{\varepsilon})$ \\\hline
    \textbf{\algWindow ($\shortWin$)$^{*}$} & sliding window & $d$-knapsack & $\frac{1-\varepsilon}{1+d}$ & $\mathcal{O}(\frac{\sqrt{W}\cdot\log(d \cdot \gamma^{-1})}{\varepsilon})$ \\\hline
    \textbf{\algWindowOpt ($\shortWinOpt$)$^{*}$} & sliding window & $d$-knapsack & $\frac{1-\varepsilon'}{2+2d}$ & $\mathcal{O}(\frac{\log(d\cdot\gamma^{-1})}{\varepsilon'}\cdot(\gamma^{-2}+\frac{\log\theta}{\varepsilon'}))$ \\\hline
  \end{tabular}
\end{table}

\textbf{Representative subset selection} (\rss) is an important tool to draw insights
from massive datasets. Existing \rss techniques can be categorized into four classes
based on the utility functions used to evaluate the representativeness:
(1) \emph{coverage-based \rss}~\cite{DBLP:conf/sdm/SahaG09,DBLP:conf/naacl/LinB10,DBLP:conf/acl/LinB11,DBLP:conf/ijcai/TamWTYH17,DBLP:conf/naacl/WeiLKB13,DBLP:conf/sigmod/XuKM14,DBLP:conf/cikm/ZhuangRHGHA16};
(2) \emph{entropy-based \rss}~\cite{DBLP:conf/kdd/BadanidiyuruMKK14,DBLP:conf/icml/GomesK10,DBLP:conf/icml/WeiIB15,DBLP:conf/www/EpastoLVZ17};
(3) \emph{clustering-based \rss}~\cite{DBLP:conf/kdd/BadanidiyuruMKK14,DBLP:conf/icml/GomesK10,DBLP:conf/nips/LindgrenWD16,DBLP:conf/uai/MalioutovKY16};
(4) \emph{diversity-aware \rss}~\cite{DBLP:conf/naacl/LinB10,DBLP:conf/acl/LinB11,DBLP:conf/icml/MirzasoleimanBK16}.
Coverage-based approaches treat \rss as the maximum coverage problem~\cite{DBLP:conf/sdm/SahaG09}
and its variants, e.g., budgeted coverage~\cite{DBLP:conf/naacl/LinB10,DBLP:conf/acl/LinB11},
weighted coverage~\cite{DBLP:conf/ijcai/TamWTYH17,DBLP:conf/cikm/ZhuangRHGHA16},
and probabilistic coverage~\cite{DBLP:conf/sigmod/XuKM14}.
They consider all information in a dataset as a collection of \emph{information units}.
The objective of \rss is to select a subset of elements so as to maximally cover the
\emph{information units} in the source dataset.
Entropy-based \rss~\cite{DBLP:conf/kdd/BadanidiyuruMKK14,DBLP:conf/icml/GomesK10,DBLP:conf/icml/WeiIB15,DBLP:conf/www/EpastoLVZ17}
(a.k.a. active set selection) aims to select a subset of elements
with the highest information entropy. Active set selection is considered as a powerful tool
for large-scale nonparametric learning~\cite{DBLP:conf/kdd/BadanidiyuruMKK14,DBLP:conf/icml/GomesK10}.
Clustering-based \rss~\cite{DBLP:conf/kdd/BadanidiyuruMKK14,DBLP:conf/icml/GomesK10,DBLP:conf/nips/LindgrenWD16,DBLP:conf/uai/MalioutovKY16}
(a.k.a. exemplar clustering) selects a subset of elements such that the average
distance from the remaining elements in the dataset to their nearest neighbor in the selected subset
is minimized. Diversity-aware \rss~\cite{DBLP:conf/naacl/LinB10,DBLP:conf/acl/LinB11,DBLP:conf/icml/MirzasoleimanBK16}
integrates a coverage/clustering based utility function
with a diversity function to avoid including highly similar elements into the selected subset.
Generally, the utility functions used in the aforementioned \rss problems are all \emph{submodular}
because the representativeness naturally satisfies the ``diminishing returns'' property.
But most of them~\cite{DBLP:conf/naacl/LinB10,DBLP:conf/acl/LinB11,DBLP:conf/naacl/WeiLKB13,DBLP:conf/sigmod/XuKM14,DBLP:conf/cikm/ZhuangRHGHA16,DBLP:conf/icml/WeiIB15,DBLP:conf/nips/LindgrenWD16,DBLP:conf/uai/MalioutovKY16,DBLP:conf/icml/MirzasoleimanBK16}
can only work in the batch setting and are very inefficient to process data streams.

Recently, we have witnessed the growth of \rss studies in the data stream model.
\rss in append-only streams where new elements arrive continuously but old ones never expire
is studied in~\cite{DBLP:conf/sdm/SahaG09,DBLP:conf/kdd/BadanidiyuruMKK14,DBLP:conf/icml/GomesK10,DBLP:conf/ijcai/TamWTYH17}.
Mirzasoleiman et al.~\cite{DBLP:conf/icml/MirzasoleimanK017} further propose a method for deletion-robust \rss
where a limited number of old elements can be deleted from the stream.
However, these techniques neither support general constraints beyond cardinality
nor consider the recency of selected subsets.
In many scenarios, data streams are highly dynamic and evolve over time.
Therefore, recent elements are more important and interesting than earlier ones.
The \emph{sliding window}~\cite{DBLP:journals/siamcomp/DatarGIM02} model is
widely adopted in many data-driven
applications~\cite{DBLP:journals/pvldb/WangZZLH16,DBLP:journals/pvldb/WangFLT17}
to capture the recency constraint.
\rss over sliding windows is still largely unexplored yet and, to the best of our knowledge,
there is only one existing method~\cite{DBLP:conf/www/EpastoLVZ17} for dynamic \rss over sliding windows.
But it is specific for the cardinality constraint.
In this paper, we propose more general frameworks for \rss than any existing ones,
which work with various submodular utility functions,
support $d$-knapsack constraints, and maintain the representatives over sliding windows.

\textbf{Submodular maximization} (\sm) has been extensively studied in recent years.
Due to its theoretical consequences, \sm is seen as a ``silver bullet'' for many different
applications~\cite{DBLP:conf/globalsip/YuXC16,DBLP:conf/kdd/LeskovecKGFVG07,
DBLP:journals/pvldb/WangFLT17,DBLP:conf/sigmod/LiFZT17,DBLP:journals/tkde/LiFWT18}.
Here, we focus on reviewing existing literature on \sm that is closely related
to our paper: \pName and \sm in data streams.
Sviridenko~\cite{DBLP:journals/orl/Sviridenko04} and Kulik et al.~\cite{DBLP:conf/soda/KulikST09}
first propose approximation algorithms for \sm subject to $1$-knapsack and $d$-knapsack constraints
respectively. Both algorithms have high-order polynomial time complexity
and are not scalable to massive datasets. More efficient algorithms for \sm
subject to $1$-knapsack constraints are proposed in~\cite{DBLP:conf/kdd/LeskovecKGFVG07,DBLP:conf/acl/LinB11}
and~\cite{DBLP:conf/soda/BadanidiyuruV14} respectively.
These algorithms cannot be applied to \pName directly.
Badanidiyuru et al.~\cite{DBLP:conf/kdd/BadanidiyuruMKK14}
and Kumar et al.~\cite{DBLP:journals/topc/KumarMVV15} propose
the algorithms for \sm with cardinality constraints
in append-only streams with sublinear time complexity.
Then, Huang et al.~\cite{DBLP:conf/approx/HuangKY17} propose an algorithm for \sm
in append-only streams with $1$-knapsack constraints.
Yu et al.~\cite{DBLP:conf/globalsip/YuXC16} propose an algorithm for \pName in append-only streams.
More recently, there are a few attempts at \sm over sliding windows.
Epasto et al.~\cite{DBLP:conf/www/EpastoLVZ17} propose an algorithm for \sm over sliding windows
with cardinality constraints.
To the best of our knowledge,
there is no existing literature on \pName over sliding windows yet.

We compare the above \sm algorithms theoretically in Table~\ref{tbl:comparison}.
We present their data models, supported constraints, approximation factors, and time complexities respectively.
According to the results, our contributions in this paper are two-fold:
(1) $\shortKs$ improves the approximation factor of \pName in append-only streams
from $\frac{1}{1+2d}-\varepsilon$ to $\frac{1-\varepsilon}{1+d}$;
(2) $\shortWin$ and $\shortWinOpt$ are among the first algorithms for \pName in the sliding window model.

\section{Conclusion}
\label{sec:conclusion}
In this paper, we studied the \emph{representative subset selection} (\rss) problem
in data streams. First of all, we formulated dynamic \rss as maximizing a monotone
submodular function subject to a $d$-knapsack constraint (\pName) over sliding windows.
We then devised the $\shortWin$ framework for this problem.
Theoretically, $\shortWin$ provided solutions for \pName
over sliding windows with an approximation factor of $\frac{1-\varepsilon}{1+d}$.
Furthermore, we proposed a more efficient $\frac{1-\varepsilon'}{2+2d}$-approximation
$\shortWinOpt$ framework for \pName over sliding windows.
The experimental results demonstrated that $\shortWin$ and $\shortWinOpt$ run orders
of magnitude faster than the batch baseline while preserving high-quality solutions.

\bibliographystyle{plainnat}
\bibliography{references}
\end{document}